\documentclass[preprint,12pt]{elsarticle}
\makeatletter
\def\ps@pprintTitle{%
 \let\@oddhead\@empty
 \let\@evenhead\@empty
 \def\@oddfoot{}%
 \let\@evenfoot\@oddfoot}
\makeatother
\usepackage{subfigure}
\usepackage{multirow}
\usepackage{graphicx}
\usepackage{amsmath}
\usepackage{natbib}
\usepackage{lineno}
\usepackage{rotating}
\usepackage[section,subsection,subsubsection]{extraplaceins}
\usepackage[colorlinks=true, linkcolor=blue, backref=page]{hyperref}%
\journal{EPJ C}
\usepackage[splitrule]{footmisc}
\usepackage{url}

\begin{document}
\begin{frontmatter}

\makeatletter
\def\@author#1{\g@addto@macro\elsauthors{\normalsize%
    \def\baselinestretch{1}%
    \upshape\authorsep#1\unskip\textsuperscript{%
      \ifx\@fnmark\@empty\else\unskip\sep\@fnmark\let\sep=,\fi
      \ifx\@corref\@empty\else\unskip\sep\@corref\let\sep=,\fi
      }%
    \def\authorsep{\unskip,\space}%
    \global\let\@fnmark\@empty
    \global\let\@corref\@empty  
    \global\let\sep\@empty}%
    \@eadauthor={#1}
}
\makeatother


\title{General formulation of process noise matrix for track fitting with Kalman filter}
\title{Analytical computation of process noise matrix in Kalman filter for fitting curved tracks in magnetic field within dense, thick scatterers}

\author
{Kolahal Bhattacharya\corref{cpa}}
\cortext[cpa]{Corresponding author}
\ead{kolahalb@gmail.com}

\author{Sudeshna Banerjee}
\author{Naba K Mondal}

\date{}


\address{Department of High Energy Physics, INO-Group (HECR Section), Tata Institute of Fundamental Research, 1-Homi Bhabha Road, Colaba, Mumbai-400005, India}
\begin{abstract}
In the context of track fitting problems by a Kalman filter, the appropriate functional forms of the 
elements of the random process noise matrix are derived for tracking through thick layers of dense 
materials and magnetic field. This work complements the form of the process noise matrix obtained by 
Mankel~\cite{mankel1997ranger}.
\end{abstract}

\end{frontmatter}
\section{Introduction}
Kalman filter~\cite{1960Kalman} is a versatile algorithm that has wide applications in various fields, 
like~\cite{smith1983monitoring,harvey1990forecasting,siouris1997tracking,vauhkonen1998kalman,blackrnan1999design,jetto1999development,mankel2004pattern,wang2005real,hoppner2010novel} etc. In 1987, 
Fr$\ddot{\rm u}$hwirth~\cite{Fruhwirth:1987fm} demonstrated its application to track fitting problems 
in high energy physics experiments for the first time. Since then, many experiments adopted this tool 
for track fitting purpose (for example,~\cite{fujiiextended,ankowski2006measurement}) and 
various authors contributed to different aspects of the algorithm (for example,~\cite{fontana2007track,Gorbunov:2006pe}). 
The problem is to estimate the charges, momenta, directions etc. of the observed particles from the 
measurements performed along their tracks. 



These parameters are combined together to form a state vector. Usually, a Kalman filter based program 
(estimator) deduces the near-optimal values of the elements of the state vector iteratively, from the 
weighted averages of the predicted locations of the particle positions 
and the measured particle positions at the sensitive detector elements. In general, the prediction is 
done based on some analytical (or numerical) solution to the equation of motion of a charged particle 
passing through a dense material and magnetic field (see Ch. 3 of~\cite{fujiiextended}, 
or~\cite{Gorbunov:2006pe}, for instance). However, the prediction represents the deterministic aspect 
of the particle motion. But the motion of the particle is also affected by the random processes like multiple 
Coulomb scattering~\cite{lynch1991approximations} and energy loss fluctuations~\cite{avdeichikov1974energy}. 
These are the stochastic perturbations to the deterministic motion of the particle, the latter being controlled 
by the magnetic field and the average energy loss. The estimator must take into account the random fluctuations 
appropriately, because precision of the filter estimation depends crucially on proper treatment of these random 
processes. Clearly, when the charged particle passes through thick layers of dense materials, the 
effects of such fluctuations are greater. 

\begin{figure}[ht]
\centering
\subfigure[Detector geometry]
{
  \includegraphics[width=0.45\textwidth,height=0.35\textwidth]{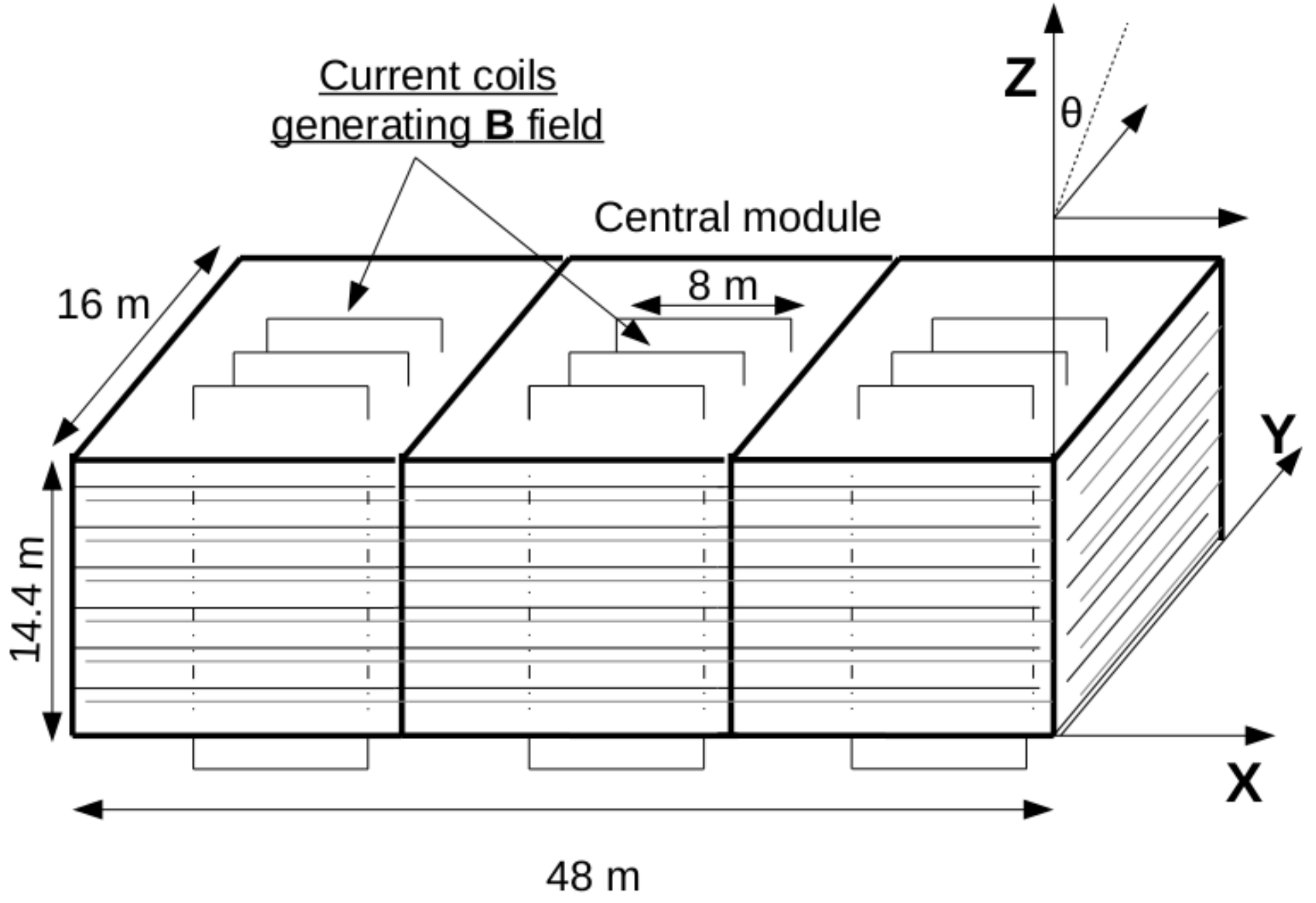}
  \label{f1(a)}
}
\hspace{0.05 cm}
\subfigure[Magnetic field map]
{
  \includegraphics[width=0.45\textwidth,height=0.35\textwidth]{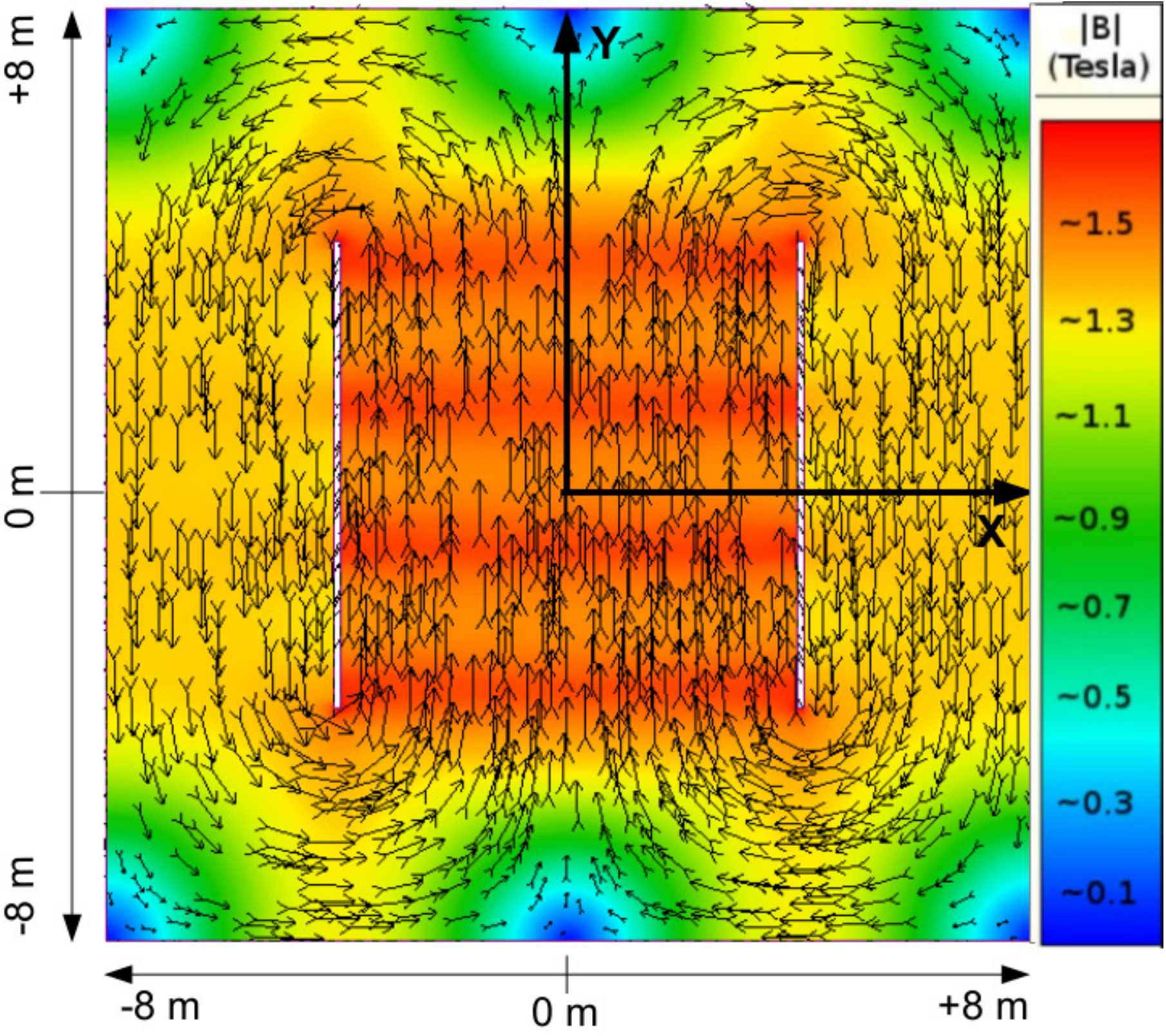}
  \label{f1(b)}
}
\caption{(a) ICAL detector geometry and (b) Magnetic field map shown in central 
module. The same field pattern exists in side modules as well. Figures taken from~\cite{bhattacharya2014error}.}
\end{figure}

This situation arises in case of track fitting in the Iron CALorimeter (ICAL) experiment, which is an 
upcoming neutrino oscillation experiment under the India-based Neutrino Observatory (INO) project~\cite{mondal2012india}. It 
comprises a 50 kiloton magnetized iron calorimeter detector of dimension $48\rm m\times 16\rm m\times 
14.4\rm m$, divided into three identical modules, as seen in Figure~\ref{f1(a)}. The sensitive detector 
elements are made by $2{\rm m}\times2{\rm m}$ resistive plate chamber detectors (RPCs), placed horizontally, 
which are sandwiched between 5.6 cm thick plates of iron. RPCs are planes 
of constant Z coordinates and any two RPCs are separated by vertical width of 9.6 cm. The iron plates 
are magnetized with current coils which generate up to $1.5\rm T$ of magnetic field (Figure~\ref{f1(b)}). ICAL 
will try to resolve the neutrino mass hierarchy, by observing the earth matter effect on the neutrino 
oscillation. The experiment is most capable of the measurement of the properties of the muons, coming 
from the charged current interactions of the muon-neutrinos. These muons travel through the different 
layers of detector materials and leave electronic signals at the RPC planes. The position measurements 
done from these signals are used for track fitting. Since ICAL will observe atmospheric neutrinos of a 
wide energy range ($E_\nu\in1 - 15\ \rm GeV$) coming from all directions, it is clear that a major 
fraction of muon tracks will be strongly affected by multiple scattering, while crossing the horizontal 
thick layers of iron at various angles. The thickness and the radiation lengths of the dense materials 
that the muons have to pass through within the ICAL detector are shown in the following Table:

\begin{table}[ht]
\small
\begin{center}
\begin{tabular}{|c|c|c|c|c|c|}
\hline
\hline
Materials     	& Iron		& RPC-glass 	& Graphite	& Copper	& Aluminium\\\hline
Thickness (cm)	& 5.6		& 0.3		& 3.00002e-03	& 9.99999e-03	& 0.0150001\\\hline
Rad. length (cm)& 1.75667	& 11.6285	& 19.2293	& 1.43516	& 8.87889  \\\hline
\hline
\end{tabular}
\end{center}
\end{table}
The width of the scattering angle is related inversely to the particle momentum and the radiation length 
of the material it is passing through~\cite{Wolin:1992ti}. So, the muons will be subjected to significant 
amount of multiple scattering inside iron. The effect will clearly be more pronounced at lower energy. It 
is also important to note that the flux of atmospheric neutrinos is much higher at lower energy~\cite{athar2013atmospheric}. 
Thus, the ICAL track fitting program must account for the random effects in a proper fashion.

Let us consider the state vector ${\bf{x}}=(x,y,t_x,t_y,q/p)^T$ which is used in many experiments like INO-ICAL~\cite{mondal2012india,ahmed2015physics,athar2013atmospheric,bhattacharya2014error}, 
MINOS~\cite{thomson2005status, lee2003description, marshall2008study}, LHCb~\cite{lhcb2008lhcb,lhcb2015measurement,lhcb2004brunel,van2005track} with forward detector geometry. Since the Kalman prediction is performed along an approximate particle trajectory, it 
introduces some deterministic uncertainties (dependent on magnetic field, average energy loss etc.) to 
the elements of the state vector. The random processes introduce additional uncertainties 
to these elements. These uncertainties are accounted for in a error covariance matrix $C=\langle({\bf{
x}}-{\bf\bar{x}})({\bf x}-{\bf\bar{x}})^T\rangle$, where ${\bf\bar x}$ contains the true values of the 
elements of the state vector. The total error matrix propagated from a point $l$ to the next $l+dl$ along the 
track is given by:

\begin{equation}\label{Eq1}
 C_{l+dl}=FC_lF^T+Q
\end{equation}
where $F$ denotes the Kalman propagator matrix, encoding the deterministic factors between $l$ and $l
+dl$. $F$ propagates the errors of the track parameters, represented by $C$ matrix, deterministically, 
from $l$ to $l+dl$. On the other hand, the matrix $Q$ represents the error contributions from all the 
random processes to the total error $C$ at $l+dl$. However, between two measurement sites, 
separated by some distance, the track fitting program should be sensitive to the possible variations of 
track parameters (momenta, direction etc.) and also to the possible variations of ambient parameters 
(materials, magnetic field components etc.). Then, one must apply Eq.\eqref{Eq1} repeatedly, in small tracking 
steps, while approaching towards the next measurement site. Thus, the effective propagator 
matrix becomes $F=\Pi_{j=1}^{N}F_j$ between two measurement sites~\cite{bhattacharya2014error}. Hence, 
the total propagated error at the next measurement site equals the sum of the (a) matrix representing 
deterministic error propagation $(\Pi_{j=1}^{N} F_j)\ C_{l_0}\ (\Pi_{j=1}^{N} F_j)^T$ and the (b) sum 
of the matrices of the {\it deterministically propagated} random uncertainties in all the tracking steps. It 
can be shown from Eq.\eqref{Eq1} that this term becomes equal to (Eq. (3.16) of~\cite{fujiiextended}):

\begin{equation}\label{Eq2}
 \sum_{m_s=1}^{N}F_{m_s,k}Q_{m_s}F_{m_s,k}^T
\end{equation}
where $F_{m_s,k}$ denotes the product of $F_j$s between $m_s$-th step and the final step. That is, to 
propagate the random uncertainties of a `deeper' layer, a longer `chain' of $F_j$s is required.

The variances of the position, angle and the momentum elements of the state vector, arising from the 
multiple scattering and energy loss fluctuation in the thin layer of dense materials, have been investigated 
by various authors~\cite{lynch1991approximations},~\cite{gluckstern1963uncertainties,billoir1984track,berger1988multiple,Wolin:1992ti,fruhwirth2001quantitative}. 
However, when passage of a particle through a thick layer of dense material is considered, one has to 
use effective variances and covariances, valid in the thick scatterer limit. These terms are obtained 
from a thorough study of Eq.\eqref{Eq2} (see Appendix B of~\cite{mankel1997ranger} written by Mankel). 
The author takes a simple form of the Kalman propagator matrix ($F$) and obtains a set of 10 ordinary 
linear coupled differential equations. The solutions to these equations correspond to the elements of 
the random noise matrix in the thick scatterer limit.

However, the result of this work is not general in two respects: (1) the propagator matrix has been assumed to be constant 
and very simple in form (see section~\ref{Sec2b}). This results in simple analytical form of elements 
of the random noise matrix $Q$ (Eq.\eqref{MankelQ}). However, in many experiments, the Kalman propagator matrix may evolve 
significantly from iteration to iteration and may have a quite non-trivial form (for example, in ICAL 
track fitting program~\cite{bhattacharya2014error}). Naturally, in these cases, one needs to find the 
more appropriate form of the random noise matrix. (2) This work~\cite{mankel1997ranger} concerns only the 
$4\times4$ block of the random noise matrix that corresponds to the position and the angular elements 
which directly suffer from multiple scattering. But the functional forms of the variance and covariance 
terms of $q/p$ with other state vector elements which are affected by the fluctuations in energy loss, 
are not considered in this work. 

The purpose of this paper is to derive the appropriate functional form of $all$ the elements of the random 
process noise matrix for a curved track in magnetic field in the thick scatterer limit. We shall take a non-trivial and evolving propagator matrix 
for this purpose and ascertain what difference it makes to the track fitting performance. Even if the 
modification does not yield significant improvements in the track fitting performance, this exercise 
serves two purposes: (a) it completes the problem from a mathematical point of view and (b) it confirms 
that Mankel's approximate solutions are good enough. To the best of knowledge of the authors, no work 
has been done before which addresses these two issues.


The problem will be formulated mathematically in the next section~\ref{Sec2}. The desired elements of 
the random noise matrix will be seen to be solutions of a matrix differential equation. Then, we will 
describe two methods of obtaining its solutions in section~\ref{Sec3}. Among these methods, the first 
one (decoupling a set of linear coupled ODEs) is practical for implementation and will be used in the 
ICAL track fitting program in the presence of magnetic field. The details of implementation technique 
will be discussed in section~\ref{Sec4}. The relevant details on software supports will be covered in~\ref{AppendC}. 
The reconstruction performance will be shown in section~\ref{Sc5}. 
We will conclude with a discussion of the merits and demerits of the approach in section~\ref{Sec5}.

\section{Mathematical formalism}\label{Sec2}
In case of the deterministic propagation of the random uncertainties, Kalman propagator matrix $F$ transports 
these uncertainties at $l$ to $l+dl$. The total random uncertainty matrix at $l+dl$ has another term coming from the 
random uncertainties introduced to the direction and the momentum of the particle due to the multiple 
scattering and the energy loss fluctuations by the material between $l$ and $l+dl$. We call this term 
$\delta Q$. The overall process noise matrix $Q$ at $l+dl$ is given by:

\begin{equation}\label{Eq3}
Q(l+dl)=F Q(l) F^T+\delta Q
\end{equation}
In Eq.\eqref{Eq3}, $F$ is the $5\times5$ propagator matrix for the Kalman filter. It can be written 
as~\cite[Eq.(24)]{fontana2007track}:

\begin{equation}\label{Eq4}
 F=I+s\ {F' dl}
  =I+s\ {
     \begin{pmatrix}
      ... & ...\\
      ... & ...
     \end{pmatrix}
      dl}
\end{equation}
where $s=+1(-1)$ when the direction of propagation increases (decreases) the $z$ coordinate while the 
tracking is carried out and $I$ denotes the identity matrix. 
The elements of the $F'$ matrix (i.e. the dots within the parenthesis of the matrix in Eq.\eqref{Eq4}) 
are the track length derivatives of the elements of the residual propagator matrix $(F-I)$. These 
quantify the additional uncertainties introduced by the presence of the magnetic field etc. to the 
existing uncertainties at $l$~\cite[pp. 10-12]{fontana2007track}. Concrete examples of the elements 
can be found from~\cite{fujiiextended},~\cite{fontana2007track},~\cite[]{bhattacharya2014error} etc. 
We shall see that the nature of

\begin{equation*}
F'\equiv
\begin{pmatrix}
... & ...\\
... & ...
\end{pmatrix} 
\end{equation*}
in Eq.\eqref{Eq4} determines the functional forms of the elements of $Q$ matrix.

\subsection{Some comments on $\delta Q$}
Since this uncertainty originates from a very small step of length $dl$, it may be assumed 
that the scattering took place in a plane of infinitesimal thickness. The elastic scattering with the 
Coulomb field of the nuclei of the dense detector material brings about a sudden change in the particle 
direction at the plane of the scattering. However, the particle position does not change laterally at 
that plane. Also, the magnitude of the momentum of the particle hardly changes as the energy imparted 
to these heavy nuclei is practically negligible~\cite[pp. 20]{mankel1995application}. If instead of $q/p$, $q/p_T$ is chosen to be a state 
element, where $p_T$ denotes the transverse momentum, it will change at that plane where the particle 
undergoes the scattering~\cite[pp. 9]{mankel1997ranger}. So, multiple scattering introduces uncertainty%
only in the particle direction and it is parametrized by two orthogonal angles $\theta_1$ and $\theta_2$, 
defined with respect to the particle direction. On the other hand, the fluctuation in the energy loss 
happens due to uncertainty in the collision rate with the atomic electron when a high energy particle 
passes through a dense material. The physical mechanism of the ionization hardly changes the particle 
direction but surely changes the magnitude of the momentum. The fluctuation, therefore, is independent 
of multiple scattering angles, but dependent on particle momentum $p$. Now, the covariance between $m
^{th}$ and $n^{th}$ elements of the state vector is given by:

\begin{equation}\label{Eq5}
 c({\bf r}_m,{\bf r}_n)=\sum_i{\frac{\partial{\bf r}_m}{\partial\xi_i}}{\frac{\partial{\bf r}_n}{\partial\xi_i}}\sigma^2(\xi_i)
\end{equation}
In Eq.\eqref{Eq5}, $\xi_i$ denotes any variable representing fluctuation due to the random processes 
(thus, $\xi=\theta_1$ or $\theta_2$ or $p$) and $\sigma(\xi_i)$ is the width of that fluctuation. 
Since $\theta_1$, $\theta_2$ and particle momentum $p$ are independent parameters, one does not need 
to calculate the covariance terms between $(\xi_i,\xi_j)$ for $i\neq j$. Then, for the chosen state vector 
$(x,y,t_x,t_y,q/p)^T$, the corresponding covariance elements may be calculated (for point scattering). 
All covariances with position coordinates $(x$ or $y)$ is zero according to our assumption that there 
is no horizontal shift of particle position in the infinitesimal plane of scattering. The covariances 
$c(t_x,q/p), c(t_y,q/p)=0$, because:

\begin{equation}\label{Eq6}
 c(t_x,q/p)=\frac{\partial t_x}{\partial\theta_1}\frac{\partial(q/p)}{\partial\theta_1}\sigma^2(\theta_1)+\frac{\partial t_x}{\partial\theta_2}\frac{\partial(q/p)}{\partial\theta_2}\sigma^2(\theta_2)+\frac{\partial t_x}{\partial p}\frac{\partial(q/p)}{\partial p}\sigma^2(p)
\end{equation}
Now, change of direction due to multiple scattering does not change $p$ and change of momentum due to 
energy loss fluctuation does not change direction $t_x$ or $t_y$. As a result, $\frac{\partial(q/p)}{
\partial\theta_{1}}=\frac{\partial(q/p)}{\partial\theta_{2}}=\frac{\partial(t_x)}{\partial{p}}=\frac{
\partial(t_y)}{\partial p}=0$. Thus, over a tracking step length $dl$, the integrated random uncertainty matrix is 
given by:

\begin{equation}\label{Eq7}
\delta Q =
 \begin{pmatrix}
       0 & 0 & 0 & 0 & 0 \\
       0 & 0 & 0 & 0 & 0 \\
       0 & 0 & c(t_x,t_x) & c(t_x,t_y) & 0 \\
       0 & 0 & c(t_y,t_x) & c(t_y,t_y) & 0 \\
       0 & 0 & 0 & 0 & c(q/p,q/p)
 \end{pmatrix}dl
\end{equation}
The nonzero variance and covariance elements of $t_x$ and $t_y$ are known in terms of the rms errors 
of the scattering angles~\cite{Wolin:1992ti,fruhwirth2001quantitative}. The calculation of $c(q/p,q/p
)$ is available from~\cite{fontana2007track}.

\subsection{Formulating the problem}\label{Sec2a}
In this section, we formulate the problem in the same way as indicated in Appendix B of~\cite{mankel1997ranger}. 
However, we also take into account the effect of energy loss fluctuation on the $q/p$ element of state vector. 
At a track length $l$, the random noise matrix $Q(l)$ is given by:

\begin{equation}\label{Eq8}
Q(l)=
 \begin{pmatrix}
  Q_{11}(l) & Q_{12}(l) & Q_{13}(l) & Q_{14}(l) & Q_{15}(l) \\
  ...       & Q_{22}(l) & Q_{23}(l) & Q_{24}(l) & Q_{25}(l) \\
  ...       & ...       & Q_{33}(l) & Q_{34}(l) & Q_{35}(l) \\
  ...       & ...       & ...       & Q_{44}(l) & Q_{45}(l) \\
  ...       & ...       & ...       & ...       & Q_{55}(l)
 \end{pmatrix}
\end{equation}
where in Eq.\eqref{Eq8}, the symmetric elements of the real symmetric matrix $Q$ has been replaced by 
dots. This shows that there are exactly fifteen independent elements of the process noise matrix that 
need to be determined. If the propagator matrix $F$ deviates from the identity matrix $I$ by a matrix 
$F'\ s\ dl$ (see Eq.~\eqref{Eq4}), then we can say:

\begin{align}\label{Eq9}
 Q(l+dl)&\approx Q(l)+Q'(l)dl\nonumber\\
        &=\left(I+F'\ s\ dl\right)\ Q(l)\ \left(I+F'\ s\ dl\right)^T+\delta Q\nonumber\\
        &\approx Q(l)+s(F'\ Q(l)+(F'\ Q(l))^T)\ dl+O(^{2})+\delta Q
\end{align}
From Eq.\eqref{Eq9}, one can deduce the differential equation of process noise $Q$:

\begin{equation}\label{Eq10}
 \frac{dQ}{dl}=s\left((F'\ Q(l)+(F'\ Q(l))^T\right)+\delta Q/dl
\end{equation}
We note that $\frac{dQ}{dl}$, $\delta{Q/dl}$ and $\left((F'\ Q(l)+(F'\ Q(l))^T\right)$ in 
Eq.\eqref{Eq10} are real symmetric matrices. This equation encodes a system of 15 coupled linear ODEs 
corresponding to the 15 independent elements of $Q$. The matrix $\left((F'\ Q(l)+(F'\ Q(l
))^T\right)$ has been calculated with the help of Mathematica~\cite{wolfram1999mathematica}, assuming all 
the elements of $F'$ are nonzero. In fact, some elements of $F'$ were found to be rather 
high (of the order of one or more) depending upon the tracking directions and momenta. The functional 
form of every element of $Q$ is the solution of the set of independent equations in Eq.\eqref{Eq10}.

\subsection{Mankel's solution}\label{Sec2b}
In his work, Mankel~\cite{mankel1997ranger} used a $4\times4$ block of random noise matrix whose 
elements were covariance terms of position and angular coordinates. The corresponding $4\times4$ 
block of the propagator matrix was given by:

\begin{equation}\label{Eq11}
F= I_{4\times4} + 
 \begin{pmatrix}
  0 & 0 & 1 & 0 \\
  0 & 0 & 0 & 1 \\
  0 & 0 & 0 & 0 \\
  0 & 0 & 0 & 0
 \end{pmatrix}
 s\ dl
\end{equation}
That is, except for $F'_{13}=F'_{24}=1$, Mankel took all the other elements of the $F'$ matrix 
to be zero. In that case, the random noise matrix has 10 independent elements. Thus, 10 linear 
coupled ODEs are obtained. The simple form of the propagator (Eq.\eqref{Eq11}) keeps the forms of the 
coupled equations simple. They can be easily solved and the resulting process noise matrix becomes: 

\begin{equation}\label{MankelQ}
Q(l)=
\begin{pmatrix}
 c(t_x,t_x)\frac{l^3}{3} & c(t_x,t_y)\frac{l^3}{3} & c(t_x,t_x)s\frac{l^2}{2} & c(t_x,t_y)s\frac{l^2}{2}\\
 ... & c(t_y,t_y)\frac{l^3}{3} & c(t_x,t_y)s\frac{l^2}{2} & c(t_y,t_y)s\frac{l^2}{2} \\
 ...    & ...    & c(t_x,t_x)l & c(t_x,t_y)l \\
 ...    & ...    & ...         & c(t_y,t_y)l
\end{pmatrix} 
\end{equation}
-where the symmetric counterparts are replaced by dots.
\section{Solution of the problem}\label{Sec3}
The matrix solution Eq.\eqref{MankelQ} is not valid in general, when all the elements of the propagator matrix are nonzero.
From Eq.\eqref{Eq10}, if the matrix connecting the fifteen independent elements of $Q$ (i.e. $Q_{11}$ 
to $Q_{55}$) to their derivatives is given as ${\bf A}_{15\times15}$, then we can write:

\begin{equation}\label{Eq12}
 \begin{pmatrix}
  \frac{dQ_{11}}{dl}\\
  \frac{dQ_{12}}{dl}\\
  ...\\
  ...\\
  \frac{dQ_{55}}{dl}
 \end{pmatrix}
 =
 s
 \begin{pmatrix}
  A_{11} & A_{12} & ... & A_{1n}\\
  A_{21} & A_{22} & ... & A_{2n}\\
  ...    & ...    & ... & ...   \\
  ...    & ...    & ... & ...   \\
  A_{n1} & A_{n2} & ... & A_{nn}
 \end{pmatrix}_{15\times15}
  \begin{pmatrix}
  Q_{11}\\
  Q_{12}\\
  ...\\
  ...\\
  Q_{55}
 \end{pmatrix}
 +
  \begin{pmatrix}
  \delta Q_{11}/dl\\
  \delta Q_{12}/dl\\
  ...\\
  ...\\
  \delta Q_{55}/dl
 \end{pmatrix}
\end{equation}
This matrix is real but not symmetric. From~\ref{AppendA}, it is seen that $110$ 
elements out of 225 elements of ${\bf A}_{15\times15}$ matrix are zero. Further simplifications arise 
from the fact that only 4 elements of the 15 elements of $\delta Q/dl$ vector are nonzero. Hence, Eq.\eqref{Eq12} 
can be succinctly written as:

\begin{equation}\label{Eq13}
 \frac{d{\bf q}}{dl}=s{\bf A}{\bf q}+\delta{\bf q}
\end{equation}
where ${\bf q}$ is a column vector of the fifteen independent elements of the $Q$ matrix $(Q_{11}, Q_{12},
...Q_{55})$ and $\delta{\bf q}$ denotes the vector of the corresponding elements of $\delta Q$ matrix 
(see~\ref{AppendA}). Within the step of length $dl$, the elements of ${\bf A}$ remain unchanged, as they 
are obtained from the propagator matrix for that step. Hence, the problem is to solve non-homogeneous 
linear coupled system of differential equations with constant coefficients. Now, we shall investigate 
different approaches for solving this initial value problem and discuss their merits and demerits.

\subsection{Solution by decoupling}\label{diag}
The most elegant method to solve Eq.\eqref{Eq13} is to decouple the equations by diagonalizing ${\bf A
}$. If ${\bf A}$ is diagonalizable (i.e. ${\bf A}=PDP^{-1}$) with an invertible $P$ and a diagonal $D
$, the system of equations can be decoupled through the substitution ${\bf q}=P{\bf u}$. In that case, 
Eq.\eqref{Eq13} reduces to:

\begin{align}\label{Eq14}
 P\frac{d{\bf u}}{dl}&=sPDP^{-1}(P{\bf u})+\delta{\bf q}\nonumber\\
 \frac{d{\bf u}}{dl} &=sD{\bf u} + P^{-1}\delta{\bf q}
\end{align}
Here $P$ is the matrix of the eigenvectors of ${\bf A}$; the corresponding eigenvalues are located at 
the diagonal position of the diagonal matrix $D$. As ${\bf A}$ is not necessarily real symmetric, the 
eigenvalues can be complex numbers as well and ${\bf A}$ may not be diagonalizable altogether in some 
cases. However, when it is diagonalizable, we can easily solve Eq.\eqref{Eq14} for ${\bf u}$ from the 
fact that the $j^{th}$ component of the equation is just a first order linear ODE: 

\begin{equation}\label{Eq15}
 \frac{du_j}{dl}=s\lambda_j u_j + (P^{-1}\delta{\bf q})_j
\end{equation}
where the set of $\lbrace\lambda_j\rbrace$ denotes the set of eigenvalues of ${\bf{A}}_{15\times15}$. 
Eq.\eqref{Eq15} can be solved by using the integrating factors and the solution to Eq.\eqref{Eq13} 
becomes:

\begin{align}\label{Eq15a}
 q_{i}(l)&=\sum_{j=1}^{15}P_{ij}u_j(l)\nonumber\\
	 &=\sum_{j=1}^{15}P_{ij}\left(e^{s\lambda_jl}u_j(0)+e^{s\lambda_jl}\int_0^l e^{-s\lambda_jl}(P^{-1}\delta{\bf q})_j\ dl\right)
\end{align}
We assume that $P^{-1}\delta{\bf q}$ varies very slowly over the small step of length $l$, so that it 
may be considered to remain constant while calculating the integral in Eq.\eqref{Eq15a}. Thus, we get:

\begin{align}\label{Eq15b}
 q_{i}(l) &\approx\sum_{j=1}^{15}P_{ij}\left(e^{s\lambda_jl}u_j(0)+e^{s\lambda_jl}(P^{-1}\delta{\bf q})_j\int_0^l e^{-s\lambda_jl}\ dl\right)\nonumber\\
          &=\sum_{j=1}^{15}P_{ij}\left[e^{s\lambda_jl}u_j(0)+e^{s\lambda_jl}(P^{-1}\delta{\bf q})_j\left(\frac{1-e^{-s\lambda_jl}}{s\lambda_j}\right)\right]\nonumber\\
          &=\sum_{j=1}^{15}P_{ij}\left[e^{s\lambda_jl}u_j(0)+\frac{(P^{-1}\delta{\bf q})_j}{s\lambda_j}(e^{s\lambda_jl}-1)\right]
\end{align}
In Eq.\eqref{Eq15b}, there are 15 unknown coefficients $u_j(0)$ that must be deduced from the initial 
conditions. The initial condition is that at $l=0$, all random noise errors are zero. We see that for 
$l=0$, Eq.\eqref{Eq15b} reduces to:

\begin{align}\label{Eq15c}
 q_{i}(0)&=\sum_{j=1}^{15} P_{ij}u_j(0)=0
\end{align}
Eq.\eqref{Eq15c} is possible only if all $u_j(0)$s are individually zero. Thus, we have:

\begin{equation}\label{Eq15d}
 q_{i}(l)=\sum_{j=1}^{15}P_{ij}\frac{(P^{-1}\delta{\bf q})_j}{s\lambda_j}(e^{s\lambda_jl}-1)
\end{equation}

In the case when ${\bf A}$ is diagonalizable, the only difficulty of implementation is the occurrence 
of complex numbers in the result. In this case, we simply take the real parts of $q_i(l)$ to form the 
elements of the random noise matrix. The imaginary parts of $q_i(l)$ cannot be used, as the imaginary 
parts of $q_1(l),q_6(l),q_{10}(l),q_{13}(l),q_{15}(l)$ (that correspond to the diagonal elements of $
Q$, i.e. the variance terms) are found to take negative values frequently. This inconsistency does 
not occur if real parts of $q_i(l)$ are used. As long as the matrix ${\bf{A}}$ is diagonalizable, $P$ 
is invertible and $\lambda_j\neq0$, this method 
is observed to work. This typically happens inside the magnetized iron plates of ICAL detector. However, 
outside iron, the conditions are not satisfied ($Det({\bf A})\rightarrow0$, one or more $\lambda_j$ 
are zero etc). As a result, Eq.\eqref{Eq15d} cannot be used there.

\subsection{Reconciliation with the process noise matrix derived in~\cite{mankel1997ranger}}
The simple process noise matrix (Eq.\eqref{MankelQ}) derived in~\cite{mankel1997ranger} is valid in a 
region with zero magnetic field where the simple form of the Kalman propagator 
matrix (Eq.~\eqref{Eq11}) is valid. On the other hand, Eq.~\eqref{Eq15d} describes the form of every 
independent element of the process noise matrix in presence of magnetic field. It is not possible to 
directly reduce $q_i(l)$ of Eq.~\eqref{Eq15d} to the corresponding elements of Eq.~\eqref{MankelQ} 
in the absence of magnetic field to check whether the generalization has been consistent, since in that 
scenario $Det({\bf A})$ becomes zero (or very close to zero) which prohibits the computations of $P
$ matrix and the eigenvalues $\lambda_j$. However, it is possible to reconcile Eq.~\eqref{Eq15d} with 
Eq.~\eqref{MankelQ} inside the magnetic field, by checking if the real parts of $q_i(l)$ are close to 
the corresponding elements in Eq.~\eqref{MankelQ}. 

Although there is an exponential dependence in Eq.~\eqref{Eq15d}, the exponent can be replaced by its 
series in the limit of small step length $l$. As a result, each term in the summation becomes a power 
law in itself and can be represented as:

\begin{align}\label{Expower}
 q_i(l)&=(P_{i1}(P^{-1}\delta{\bf q})_1 + P_{i2}(P^{-1}\delta{\bf q})_2 + ...) l \\ \nonumber
       &+(P_{i1}(P^{-1}\delta{\bf q})_1\frac{s\lambda_1}{2} + P_{i2}(P^{-1}\delta{\bf q})_2\frac{s\lambda_2}{2} + ...) l^2\\ \nonumber
       &+(P_{i1}(P^{-1}\delta{\bf q})_1\frac{(s\lambda_1)^2}{6} + P_{i2}(P^{-1}\delta{\bf q})_2\frac{(s\lambda_2)^2}{6} + ...) l^3 \\ \nonumber
       &+ ...
\end{align}
So, one needs to check if the real parts of the coefficients of $l$, $l^2$ and $l^3$ in Eq.\eqref{Expower} 
are close to the corresponding coefficients in Eq.\eqref{MankelQ}. This exercise has been performed and 
the results are shown in~\ref{AppendB}.

\subsection{Solution method without diagonalization}
In this case, one first needs to solve the homogeneous equation $
\frac{d{\bf q}}{dl}={\bf Aq}$ (where the constant factor $s$ is absorbed within the matrix ${\bf A}$). 
The solutions for the vector ${\bf q}(l)$ are used to form a fundamental matrix solution $M(l)$~\cite{kelley2010theory}, each 
column of which is independent and satisfies the homogeneous part of Eq.\eqref{Eq13}. Using the method 
of variation of parameters, the solution to the non-homogeneous initial value problem:

\begin{equation}\label{Eq16}
 \frac{d\bf q}{dl}={\bf A\ q}+\delta{\bf q},\hspace{1.0cm}{\bf q}(l_0)={\bf q}_0
\end{equation}
can be given by~\cite{kelley2010theory}:

\begin{align}\label{Eq17}
 {\bf q}(l) &=M(l)M(l_0)^{-1}{\bf q}_0+M(l)\int_{l_0}^{l}M(t)^{-1}\delta{\bf q}(t)dt
\end{align}
When it is possible to find out all the possible eigenvalues and independent eigenvectors of ${\bf A}
$, construction of $M(l)$ is straightforward~\cite[Ch.37]{KennethHowell}. However, matrices are not always diagonalizable. So, it 
is essential to have an alternative method of deriving $M(l)$ when the calculation of all independent 
eigenvectors is not possible. This can be achieved by Putzer's algorithm~\cite{kelley2010theory}. The 
method is elegant in the sense that it does not require all the eigenvalues to be distinct or nonzero. 
However, in case of solving Eq.\eqref{Eq13} it is seen that the calculation of $M(l)$, a $15\times15
$ matrix, becomes impractically lengthy, and therefore, the method has not been adopted. But if it is 
possible compute $M(l)$ by this method, that may be used even outside magnetized iron plates.

\section{Application to ICAL}\label{Sec4}
In the track fitting program for ICAL~\cite{bhattacharya2014error}, thick scatterer approximation has 
been used previously by implementing Mankel's form of random noise matrix~\cite{mankel1997ranger}. Strictly speaking, this 
form of matrix is valid only if the track segment is linear, since the magnetic field dependent terms 
(that lead to curvature of the track) are assumed to be zero in the propagator matrix $F$ (Eq.\eqref{Eq11}). 
Therefore, it is a matter of interest to see how the performance of track fitting is affected, when a 
more appropriate solution (Eq.\eqref{Eq15d}) is applied to construct the random process noise matrix 
in the presence of the magnetic field. 

This has been carried out through the use of a C++ based computational library it++~\cite{cristeait++}. 
Details of the coding techniques etc. are given in~\ref{AppendC}. It was seen that in all the 
cases where all the elements of ${\bf A}$ are non-trivial (which commonly happens within the magnetic 
field), the determinants of ${\bf A}$ assume large values $(10^0-10^6)$ and the diagonalizations can be carried out 
quite easily. However, in the regions where the magnetic field is zero (outside the iron slabs in the 
ICAL detector) or its spatial derivatives are zero (inside iron), $|Det({\bf A})|=0$ (or $|Det({\bf A})|\rightarrow0$) and Eq.\eqref{Eq15d} 
cannot be applied. This can be understood in the following way: outside iron, the propagator matrix 
reduces to Eq.\eqref{Eq11}, as all the magnetic field integrals vanish. Even inside iron, certain 
elements in the first two columns of $F'$ matrix (e.g. $F'_{11}$, $F'_{12}$ etc. which depend on 
spatial derivatives of magnetic field components~\cite{bhattacharya2014error}) become zero occasionally. These 
zeros lead to additional zeros in the matrix ${\bf A}$ and the determinant of the latter becomes very 
small (close to zero)\footnotemark[1]. That the determinant is zero (or close to zero) suggests 
\footnotetext[1]{The $F'$ matrix used by Mankel (Eq.\eqref{Eq11}) has 
determinant exactly zero. In fact, this is a limiting case, where all the elements of this matrix are 
zero except $F'_{13}=F'_{24}=1$.} 
that one or more eigenvalues are zero (or close to zero). Hence, Eq.\eqref{Eq15d} cannot be evaluated 
properly and unphysical solutions are obtained if Eq.\eqref{Eq15d} is applied. Therefore, outside 
the iron plates (and occasionally inside the iron plates) where $|Det({\bf A})|$ is small $(\le1)$, we 
used Mankel's form of the process noise matrix Eq.~\eqref{MankelQ}. In general, inside the magnetic field, 
where $Det({\bf A})$ is typically $\gg1$, we applied Eq.\eqref{Eq15d} to construct the elements of the 
process noise matrix by diagonalizing ${\bf A}$ through it++.

Since the solutions $q_i(l)$s represent the terms of a covariance matrix, we expect that $q_1$, $q_6
$, $q_{10}$, $q_{13}$, $q_{15}$ will be positive, because they correspond to the diagonal elements of 
the $Q$ matrix ($Q_{11}, Q_{22}, Q_{33}, Q_{44},Q_{55}$ respectively). However, the real parts of the 
solutions $q_i(l)$s need not be positive. It is interesting to see that the computation automatically 
led to positive values of diagonal elements, as expected. No additional measure was needed to obtain 
these positive values. This shows that the analysis has been consistent.



\section{Reconstruction performance}\label{Sc5}
Since the method of computing the process noise matrix described in this paper is somewhat abstract, 
first we would like to show that the resulting Kalman filter works in a consistent fashion. Once that 
is done, we shall check if the Kalman filter, equipped with the random noise matrix developed in this 
paper, has better (or worse!) reconstruction performance compared to the one equipped with the random 
noise matrix derived by Mankel~\cite{mankel1997ranger}.

We used GEANT4~\cite{Agostinelli2003250} to generate 5000 Monte Carlo muons ($\mu^-$) inside the ICAL 
detector. The event vertices were smeared uniformly across a volume of $(43.2\rm{ m}\times14.4\rm{ m}
\times10.0\rm{ m})$ around the center of the detector (see Figure~\ref{f1(a)}) in all $\phi$ directions ($\phi\in[0,2\pi]$). This ensures that the 
muon tracks from the inhomogeneous magnetic field region (Figure~\ref{f1(b)}) are also present in the 
total set of simulated tracks, in the same way it would happen in reality. 

\begin{figure}[ht]
\centering
\subfigure[Pull of X]
{
  \includegraphics[width=0.46\textwidth,height=0.315\textwidth]{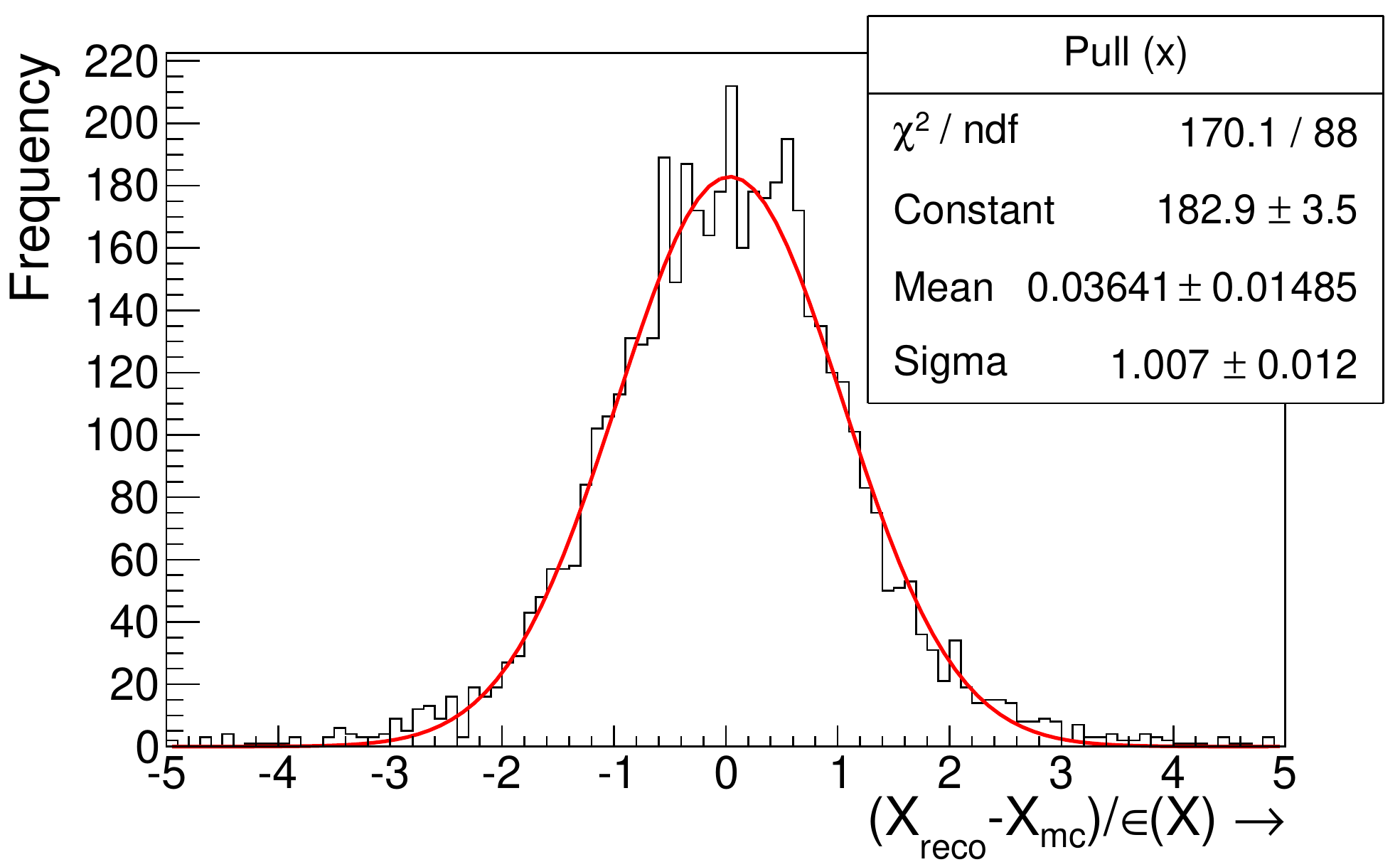}
  \label{f2:a}
}
\hspace{0.01 cm}
\subfigure[Pull of Y]
{
  \includegraphics[width=0.46\textwidth,height=0.315\textwidth]{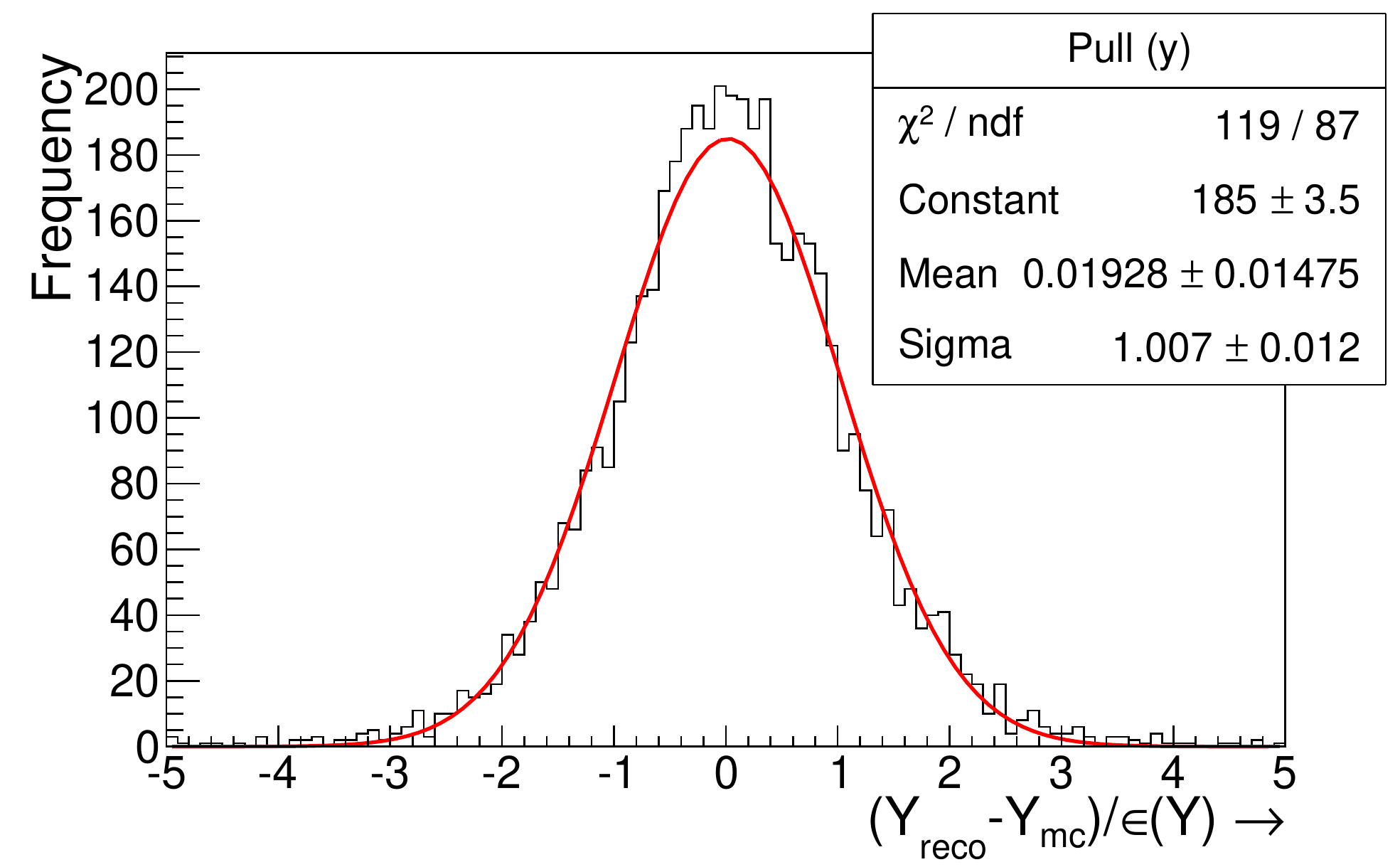}
  \label{f2:b}
}
\subfigure[Pull of $t_x$]
{
  \includegraphics[width=0.46\textwidth,height=0.315\textwidth]{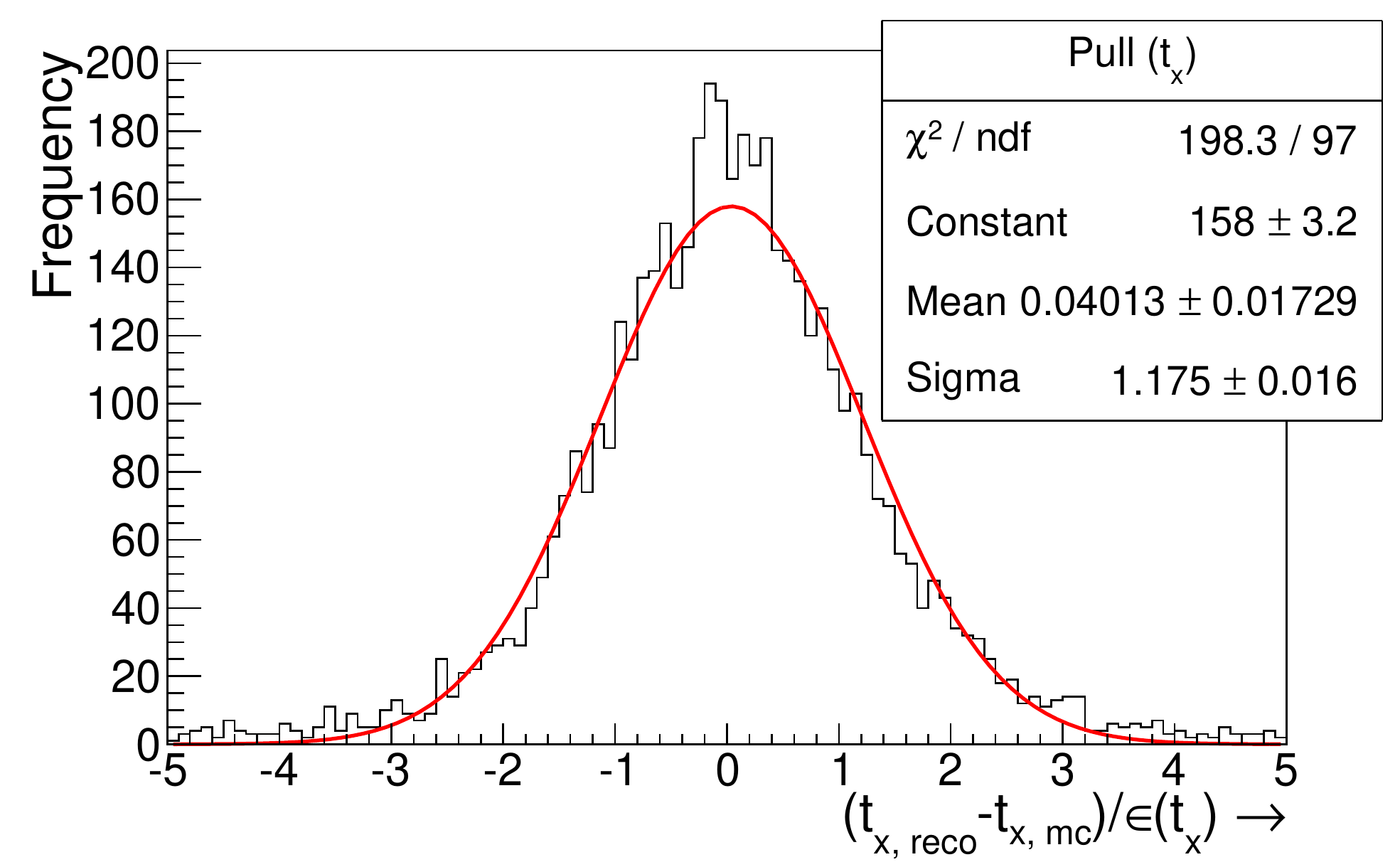}
  \label{f2:c}
}
\hspace{0.01 cm}
\subfigure[Pull of $t_y$]
{
  \includegraphics[width=0.46\textwidth,height=0.315\textwidth]{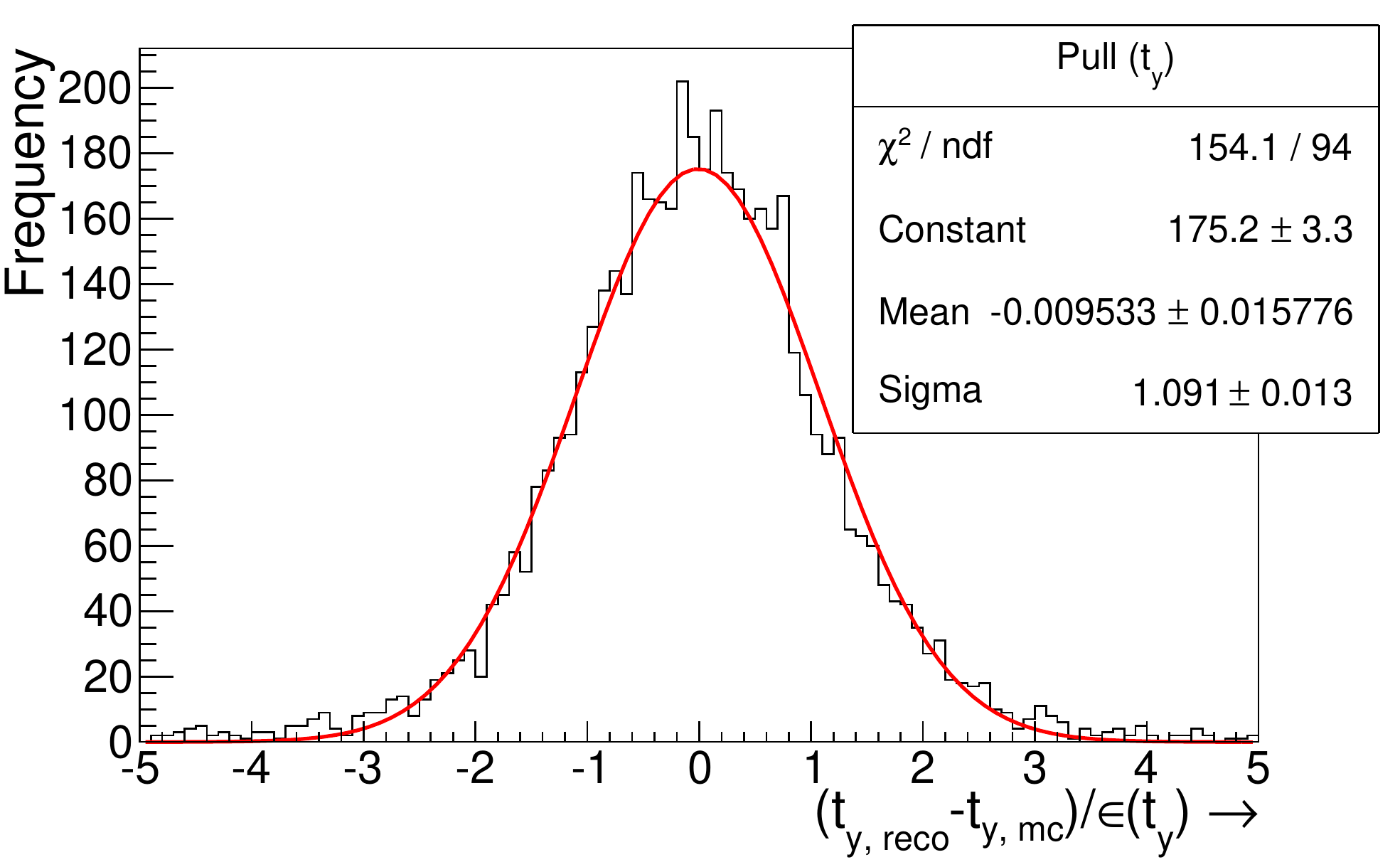}
  \label{f2:d}
}
\subfigure[Pull of $q/p$]
{
  \includegraphics[width=0.46\textwidth,height=0.315\textwidth]{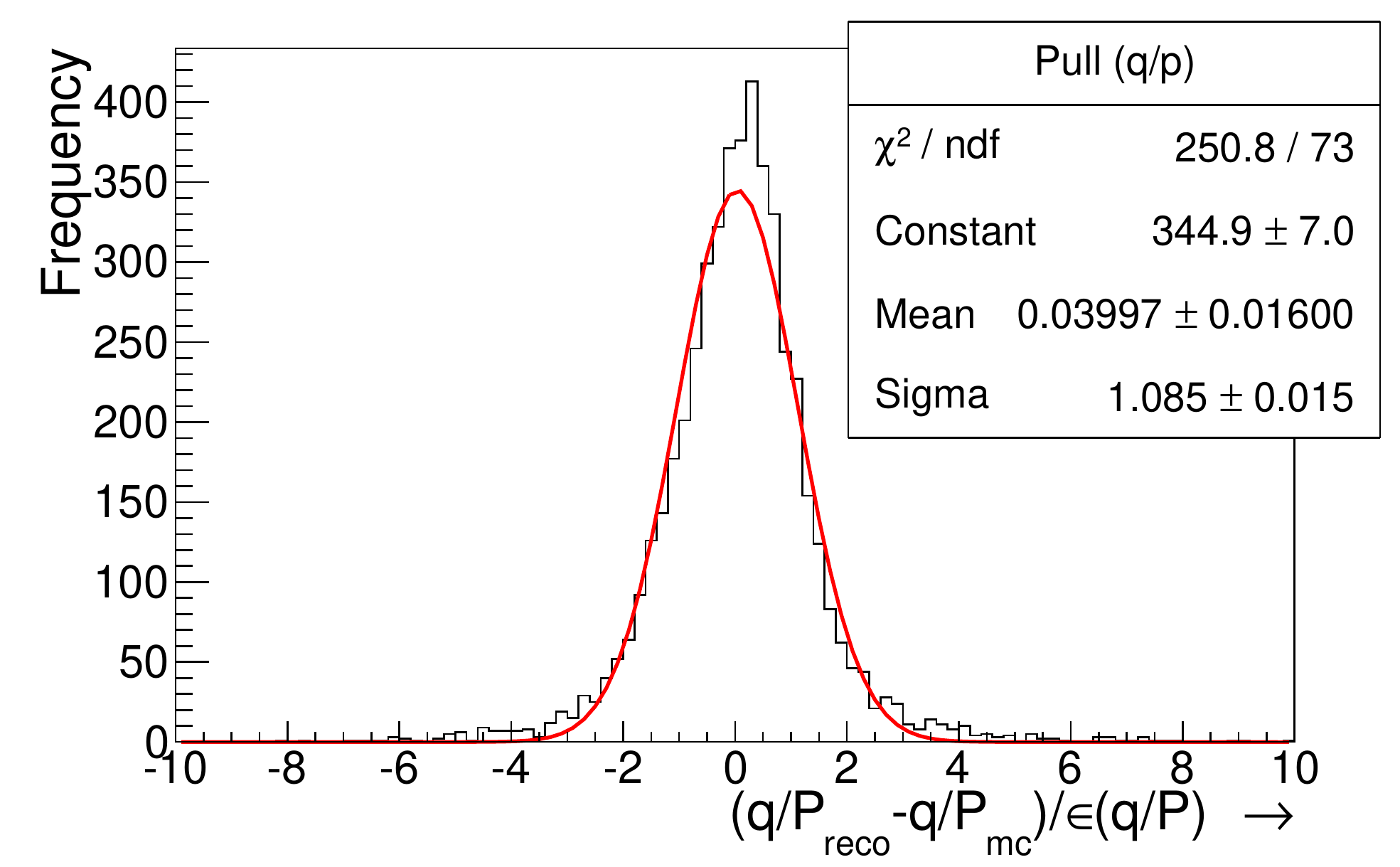}
  \label{f2:e}
}
\hspace{0.01 cm}
\subfigure[Reduced $\chi^2$]
{
  \includegraphics[width=0.46\textwidth,height=0.315\textwidth]{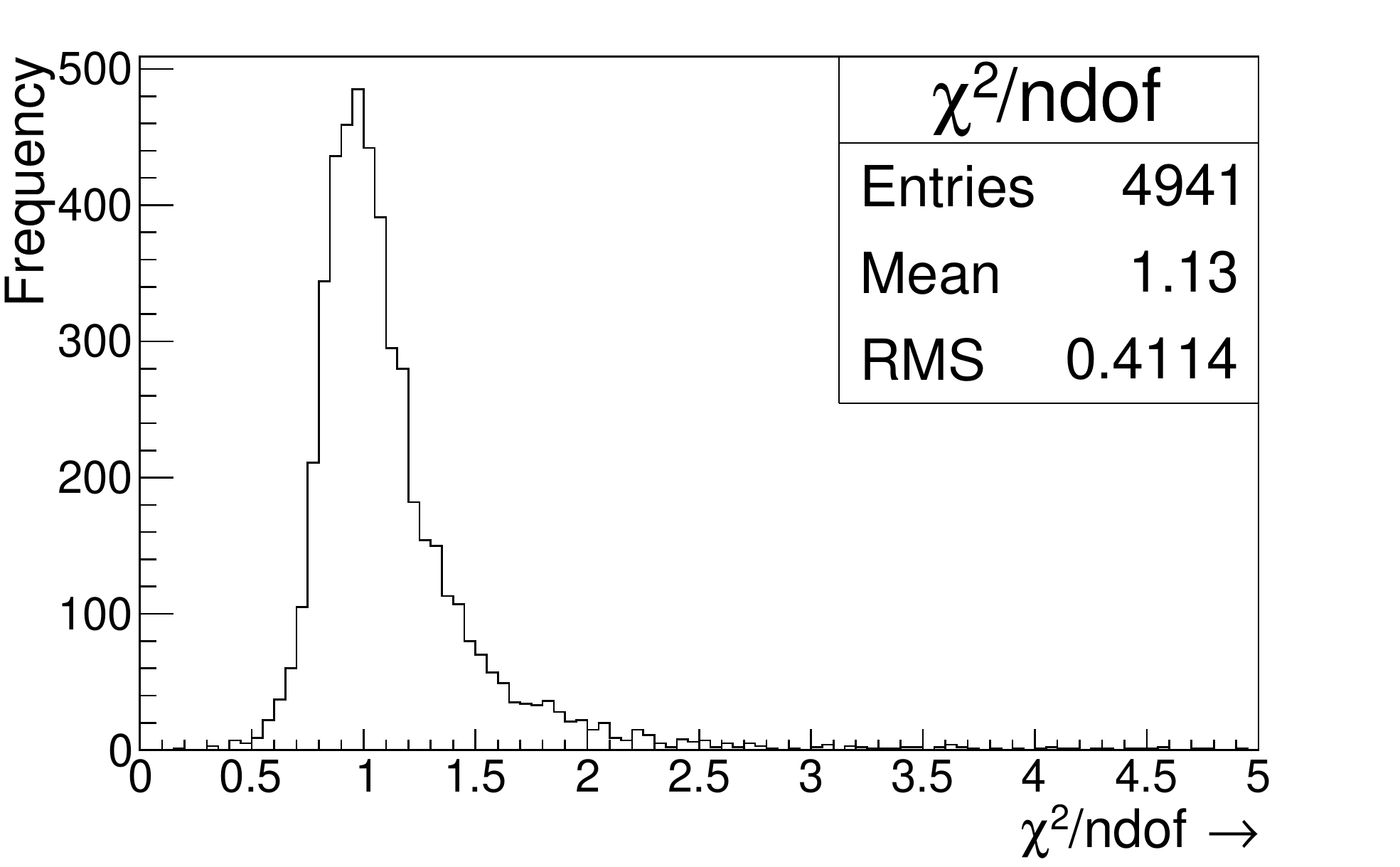}
  \label{f2:f}
}
\caption{Reconstructed muon of momentum 5 GeV/c at zenith angle $\theta=\cos^{-1}0.95$. 
(a) Pull of X, (b) pull of Y, (c) pull of $t_x$, (d) pull of $t_y$, (e) pull of $\frac
{q}{p}$ and (f) Reduced $\chi^2$. The error of a variable $\zeta$ has been denoted by 
$\epsilon(\zeta)$.}
\label{f2}
\end{figure}

To show that the filter is working in the expected way, we shall present the `goodness of fit' plots 
in the following. These are the pull distributions of the fitted variables and the reduced $\chi^2$ 
distribution. The pull of a given variable $\zeta$ is defined as:
\begin{equation}
 Pull(\zeta)=\frac{\zeta_{Reconstructed}-\zeta_{Monte\ Carlo}}{\sqrt{C_{\zeta\zeta}}}
\end{equation}
where $C_{\zeta\zeta}$ denotes the error of the reconstructed $\zeta$ parameter, as estimated from the updated covariance 
matrix of the Kalman filter. In ICAL, we are mostly interested in the fitted parameters near the muon 
event vertex; hence, the pull is evaluated there only. For good fit, the pull distributions must have 
mean at zero and standard deviation equal to unity. In Figure~\ref{f2}, we show these plots for muons 
of momentum 5 GeV/c, with initial direction $\theta=\cos^{-1}0.95$ to the vertical. 

\begin{figure}[ht]
\centering
\subfigure[Momentum Resolution]
{
  \includegraphics[width=0.46\textwidth,height=0.35\textwidth]{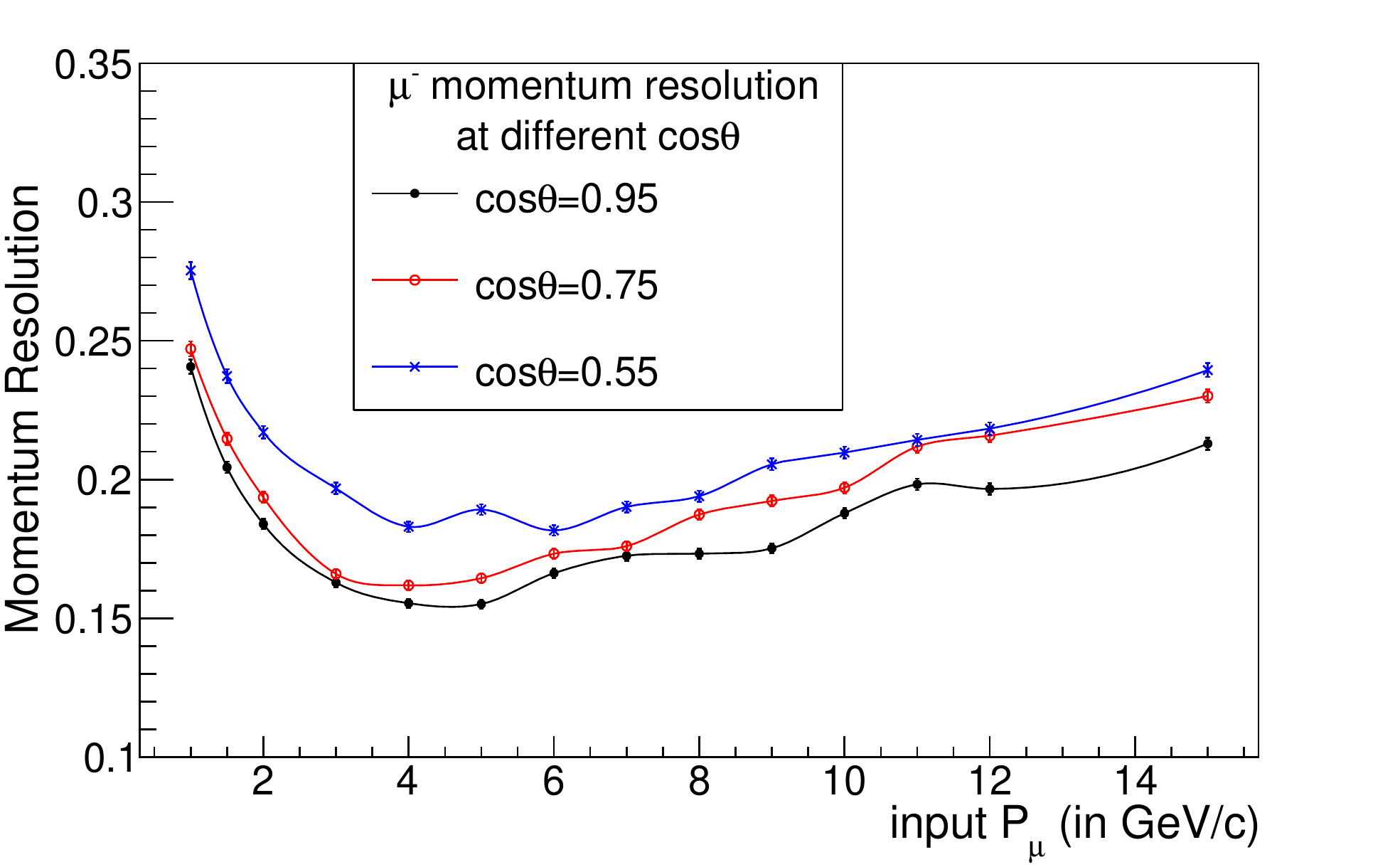}
  \label{f3:aMR}
}
\hspace{0.0 cm}
\subfigure[$\cos\theta$ Resolution]
{
  \includegraphics[width=0.46\textwidth,height=0.35\textwidth]{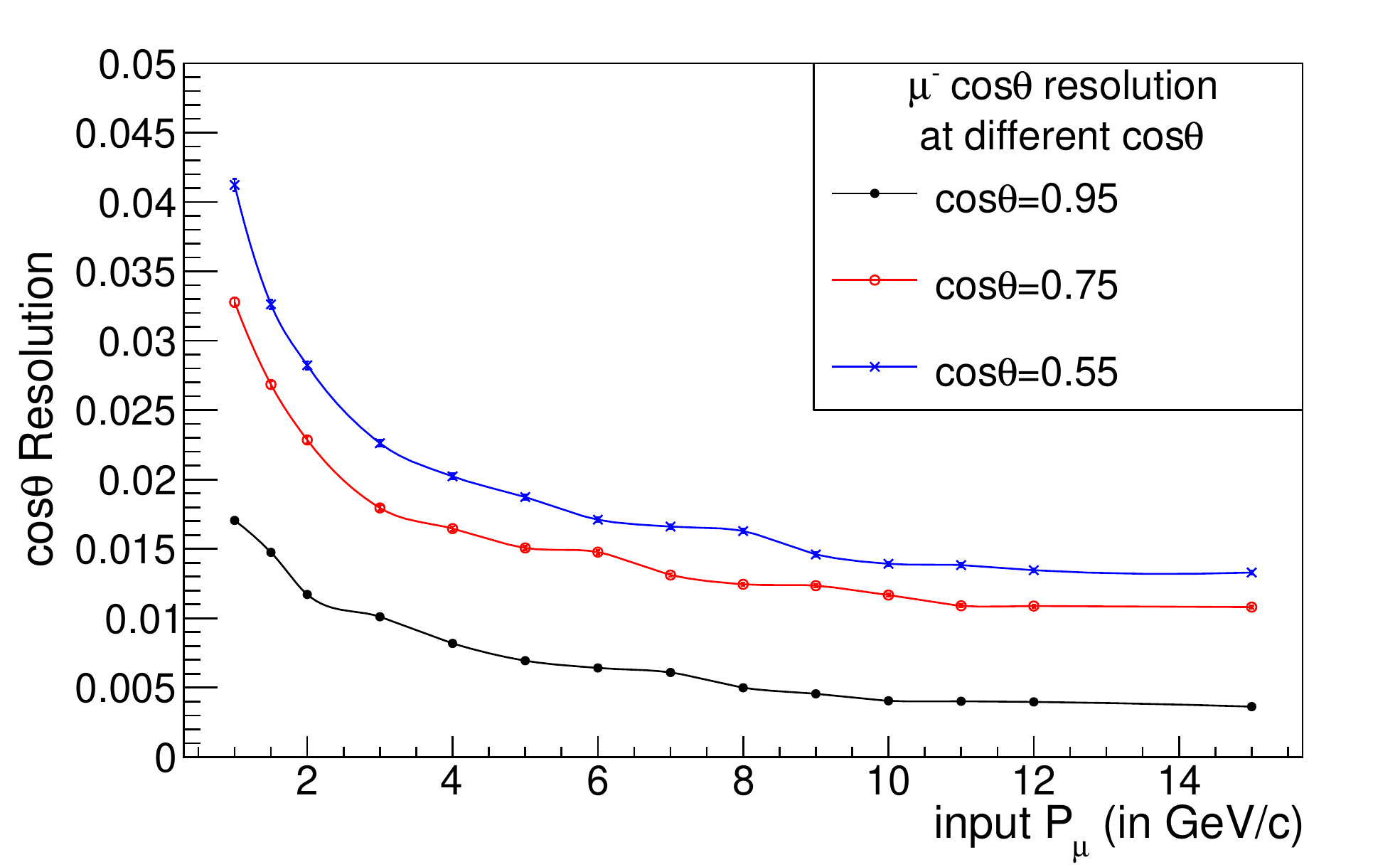}
  \label{f3:b}
}
\caption{Reconstructed (a) momentum and (b) direction ($\cos\theta$) resolution plots of muon $\mu^-$ as functions 
of increasing input momenta at various initial angles.}
\label{f3}
\end{figure}

The reduced $\chi
^2$ of the model prediction of every event is obtained by dividing the total $\chi_p^2=\sum_k{({\bf{r
}}_k^{k-1})^T(R_k^{k-1})^{-1}{\bf{r}}_k^{k-1}}$~\cite{Fruhwirth:1987fm} by the no. of free parameters. 
Here ${\bf{r}}_k^{k-1}$ denotes the residual of state prediction and $R_k^{k-1}$ is the corresponding error 
covariance matrix. The total no. of free parameters equals $2n-5$, found by subtracting 5 constraints 
(through initialization of the filter) from the total degrees of freedom (two times the no. of hits $
n$ along the track). The $\chi_p^2$ for prediction is equal to the $\chi_f^2$~\cite{Fruhwirth:1987fm} 
for the track fit. From Figure~\ref{f2}, it is observed that the pull distributions of the elements of 
the state vector have mean very close to zero and fitted width very close to unity. The reduced $\chi
^2$ plot (Figure~\ref{f2:f}) peaks close to unity as well, as expected. Similar performance of track 
reconstruction is observed in a wide range of $p$ and $\cos\theta$. 
\FloatBarrier
\FloatBarrier
At very low momenta ($p<2$ GeV/c) 
and very large angles $\theta>60^0$, gradual worsening of the reconstruction performance is observed. 
This degradation is intrinsic to the tracking problem, irrespective of whether or not the enhanced track scatterer treatment, described in this paper, is included.
The gradual worsening is seen from the following momentum and direction resolution plots in Figure~\ref{f3}. Here, the 
momentum resolution has been defined as $\frac{\sigma(p)}{p_{in}}$, where $\sigma(p)$ denotes the rms width 
of the reconstructed momentum distribution and $p_{in}$ denotes the input momentum. On the other hand, 
rms width of the reconstructed $\cos\theta$ distribution (i.e. $\sigma(\cos\theta)$) has been chosen as the 
definition of the direction resolution. 

At lower momenta and/or larger zenith angles, the muon tracks are affected by the multiple scattering 
to a greater extent. This affects the precision of muon momentum estimation, for which the resolution 
becomes poor (Figure~\ref{f3:aMR}). On the other hand, at higher muon input momenta, the contribution 
to the momentum resolution from the spatial resolution component is higher~\cite[Eq. (3.5)]{hauptman2011particle} 
and that leads to gradual worsening of momentum resolution. 
The direction resolution steadily improves with increasing $p_{in}$, but worsens as input $\theta$ is 
increased. 

Let us now proceed to the comparative study of the Kalman filters equipped with two different process 
noise matrices: one derived in this paper and the other derived in~\cite{mankel1997ranger}. To see if 
the former has any advantage or disadvantage over the latter, these two programs were used to fit the 
two copies of the simulated muon tracks of momentum $5$ GeV/c and initial direction $\theta=\cos^{-1}
0.95$. Thus, the two filters with the two different process noise matrices operated on identical sets 
of position measurements. It was observed that the quality and performance of reconstruction of these 
two programs are of the same order. For individual events, the correction in the reconstructed values 
of momentum or $\cos\theta$ usually appeared at the second or third decimal places or beyond that. In 
fact, no significant improvement or deterioration was observed for any of the track parameters. Thus, 
no gross improvement was achieved by using the more appropriate form of random process noise 
matrix inside the iron plate equipped with magnetic field. This is shown in the following Figure~\ref{f4:a} and~\ref{f4:b}.

\begin{figure}[ht]
\centering
\subfigure[]
{
  \includegraphics[width=0.46\textwidth,height=0.32\textwidth]{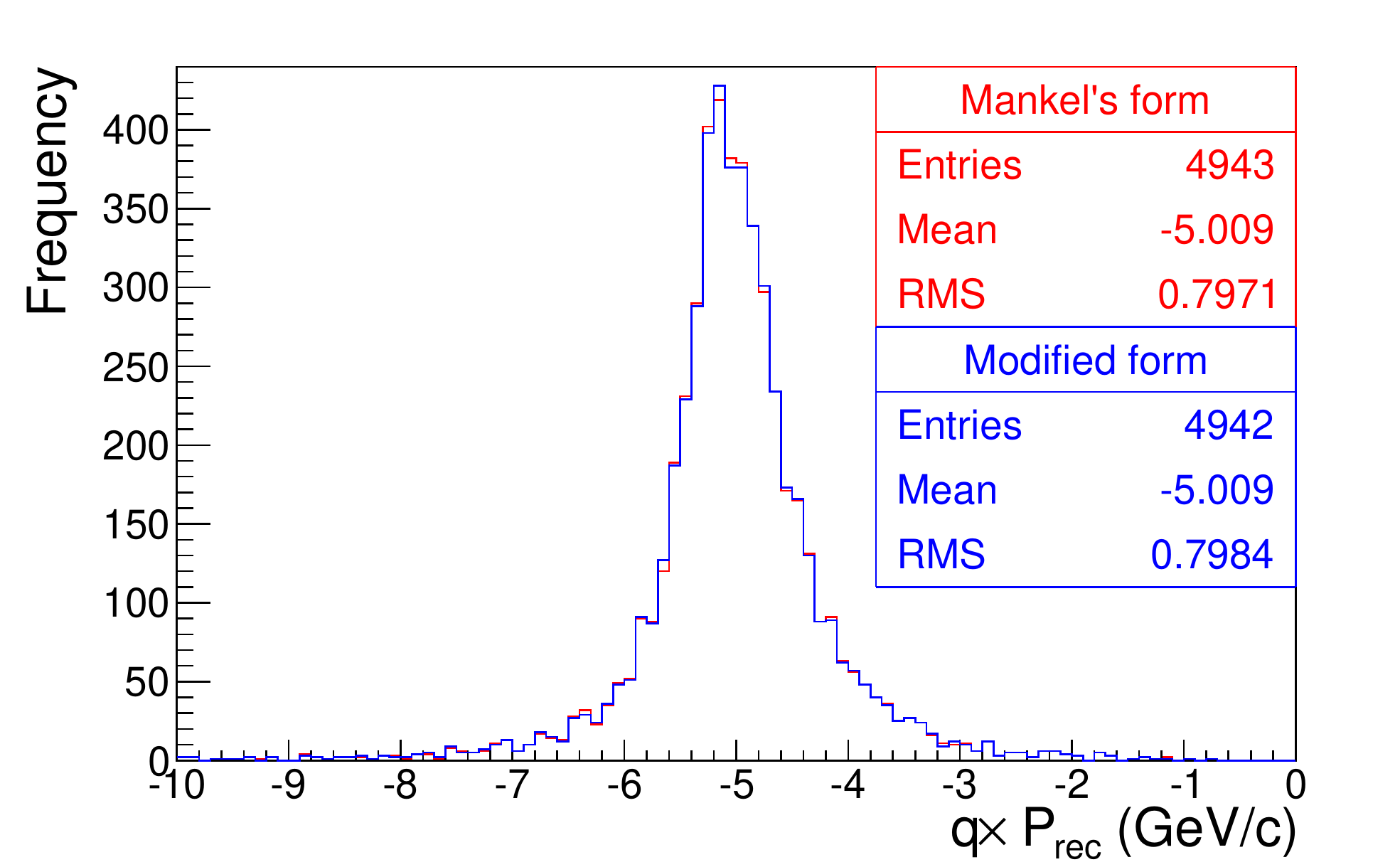}
  \label{f4:a}
}
\subfigure[]
{
  \includegraphics[width=0.46\textwidth,height=0.32\textwidth]{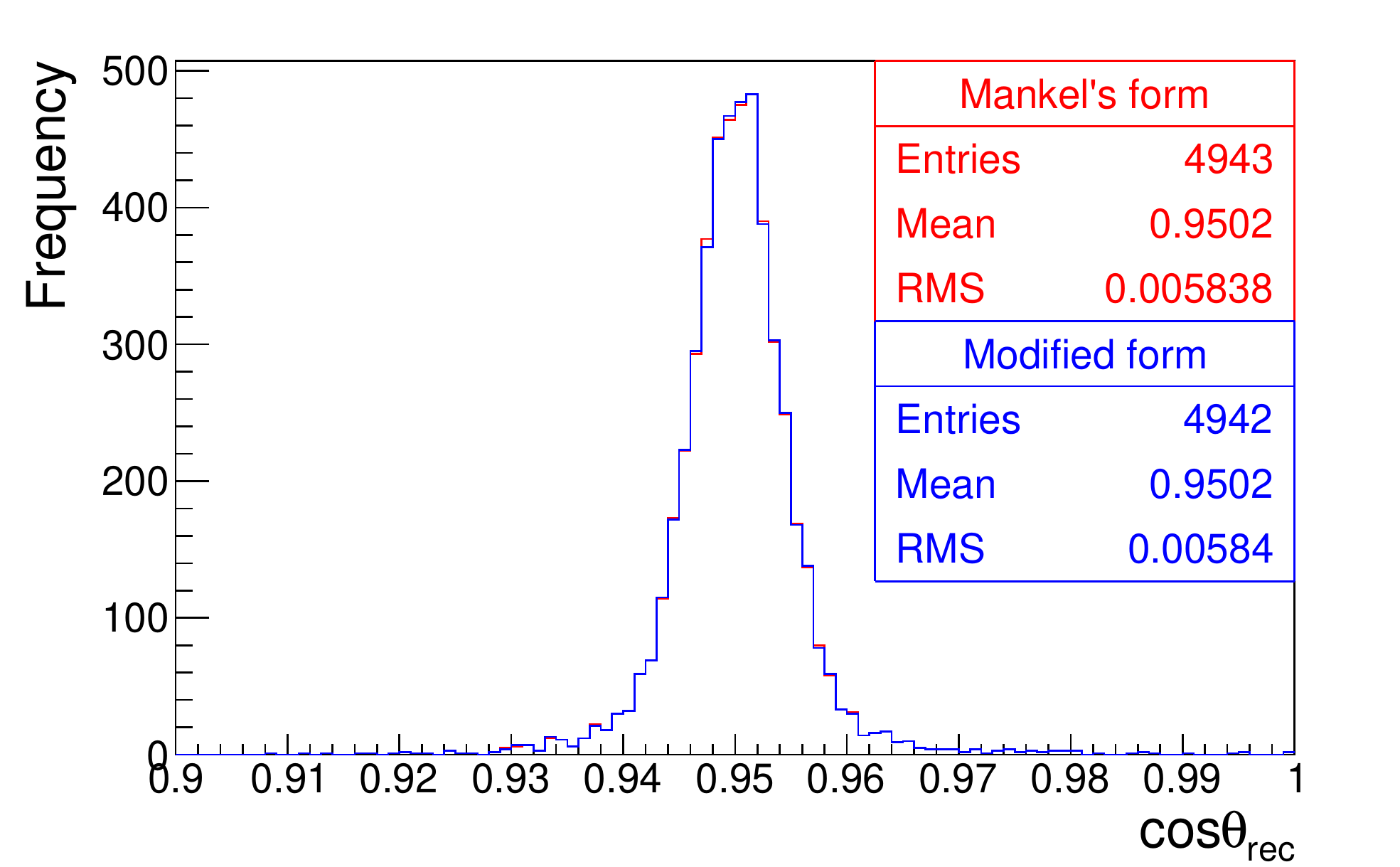}
  \label{f4:b}
}
\caption{Comparison of the track fitting performance between the Kalman filters equipped with the 
process noise matrix derived in~\cite{mankel1997ranger} and that derived in this paper. (a) Comparison 
of reconstructed momentum and (b) direction ($\cos\theta$). The result is for 5000 muon tracks of true 
momentum 5 GeV/c in the ICAL detector.}
\label{f4}
\end{figure}
\begin{figure}[ht]
\centering
\subfigure[]
{
  \includegraphics[width=0.46\textwidth,height=0.32\textwidth]{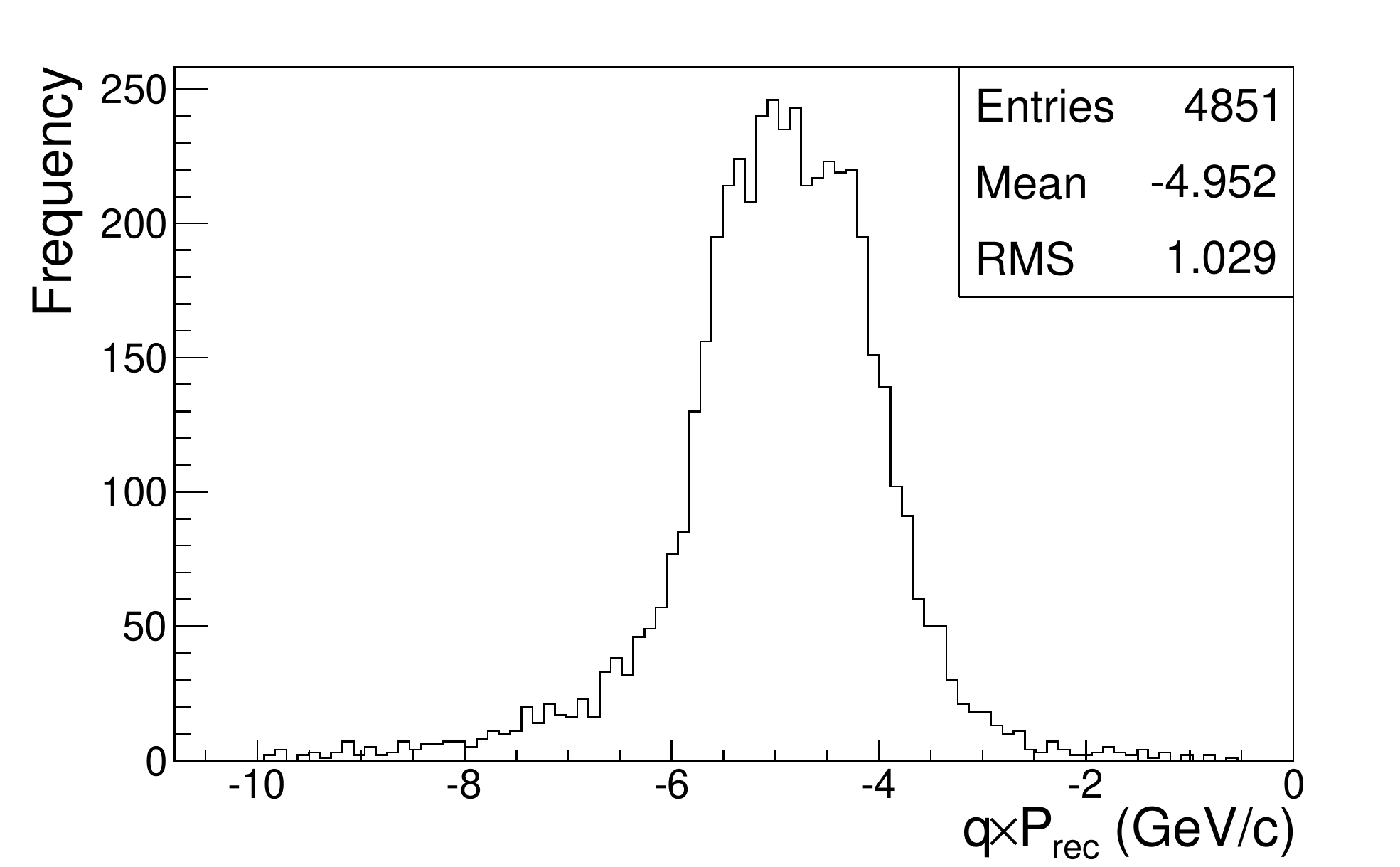}
  \label{f5:a}
}
\subfigure[]
{
  \includegraphics[width=0.46\textwidth,height=0.32\textwidth]{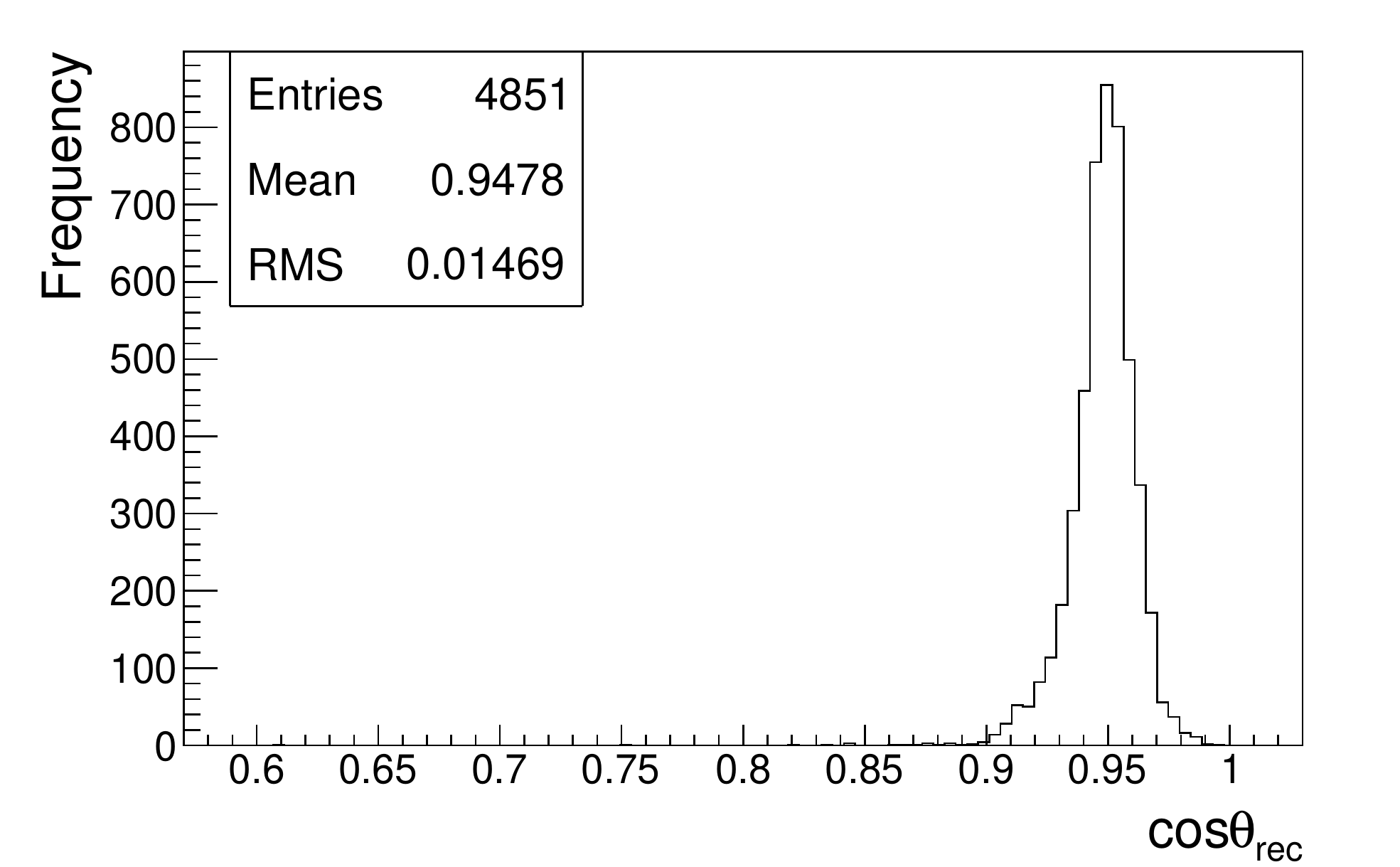}
  \label{f5:b}
}
\caption{Reconstruction performance: (a) momentum and (b) direction ($\cos\theta$) of the events with 
the elements of the random process noise matrix set to zero.}
\label{f5}
\end{figure}

This observation that there is hardly any difference in the reconstruction performance may raise some 
doubt about the validity of the process noise treatment. One may be interested to check how large the 
effect of the process noise treatment is in the first place. If the fitting is performed by switching 
off the process noise, we expect the fitting performance to deteriorate. This exercise was performed 
by setting all the elements of the process noise matrix to zero, but keeping the other of the program 
the same as before. The result is shown in Figure~\ref{f5}. It is seen that the fitting performance 
of the same set of events becomes less precise, due to inconsideration of the process noise treatment. 
In fact, many of the events are reconstructed with worse momenta values, as seen from the event count 
in Figure~\ref{f5:a} (less compared to those in Figure~\ref{f4:a}). The filter, however, converges to 
more or less accurate mean value, since it took into account the mean energy loss in correct manner. 
On the other hand, the direction estimation becomes very poor, as seen from the width of the distribution 
in Figure~\ref{f5:b}.

This consistency check also confirms that the fitting performance improves significantly with respect 
to ``no process noise treatment'', when the process noise matrix is accounted for. Figure~\ref{f4:a} 
shows that the formula of the process noise matrix developed in this paper, which was used for tracking inside magnetized iron plates, does not lead 
to gross improvement of track fitting performance. So, we conclude that Mankel's simple solution for 
random noise matrix is indeed a good approximation.

\section{Summary}\label{Sec5}
In this paper, a mathematical formalism has been developed for expressing the elements of the 
random noise matrix while performing track fitting with a Kalman filter through a thick scatterer and 
nonzero magnetic field. In this case, all the elements of the propagator are nonzero, unlike Mankel's 
approach~\cite{mankel1997ranger} and we made use of the method of diagonalization (see~\ref{diag}) to 
construct the desired elements under such circumstances. Through this formalism, the elements $\rm cov
({\bf{x}}, q/p)=\rm cov(q/p, {\bf{x}})$ of the random noise matrix can also be calculated for a track 
deflected by a magnetic field in a thick scatterer. Evaluation of these elements was not included in 
Mankel's treatment~\cite{mankel1997ranger}. Although no precaution was taken to render the real parts 
of $q_1, q_6, q_{10}, q_{13}$ and $q_{15}$ positive (which correspond to the diagonal elements of the 
random noise matrix), they turned out to be positive in all the cases. However, this solution could not 
be used outside magnetic field region. Also, its use inside the magnetized iron plates did not improve 
the track fitting performance. The treatment 
by Mankel~\cite{mankel1997ranger}, derived under approximations, seems good enough for reconstruction 
of momentum, at least to the first or second decimal place. This is also clear from Table~\ref{Table} 
in~\ref{AppendB} which shows that the corrections introduced to the elements of the process noise matrix 
are small. On the other hand, the mathematical form of the elements of the process noise matrix, 
derived in this paper, is quite general and can be used in the context of any state vector in other HEP 
experiments employing different state vectors (for example, those containing $q/p_T$ or curvature $
\kappa$ of the track as one of the elements).

\section{Acknowledgements}
K.B. expresses gratitude to the INO Cell, TIFR for providing short term visiting fellowship for supporting 
this research. The authors also acknowledge the comments of the anonymous referee for suggesting many 
improvements in the organization of the paper.

\appendix
\section{}\label{AppendA}
Let us first formally define the vector ${\bf q}$ as:
\begin{equation*}
 {\bf q}=\left(Q_{11},Q_{12},Q_{13},Q_{14},Q_{15},Q_{22},Q_{23},Q_{24},Q_{25},Q_{33},Q_{34},Q_{35},Q_{44},Q_{45},Q_{55}\right)^T
\end{equation*}
Hence, $q_1, q_6, q_{10}, q_{13}$ and $q_{15}$ represent the diagonal elements of $Q$ matrix. Then, we 
formally define the vector $\delta{\bf q}$ as (see Eq.~\eqref{Eq7}, Eq.~\eqref{Eq12}, Eq.~\eqref{Eq13}):\\
\begin{equation*}
 \delta{\bf q}=\left(0,0,0,0,0,0,0,0,0,c(t_x,t_x),c(t_x,t_y),0,c(t_y,t_y),0,c(q/p,q/p)\right)^T
\end{equation*}

Next, we construct the matrix ${\bf A}$ of Eq.\eqref{Eq12}. This is a ${15\times15}$ matrix with many 
non-trivial elements. Hence, we express it by dividing it into two blocks $B
^1$ and $B^2$ such that ${\bf A}=({\bf B}^1_{15\times8}|{\bf{B}}^2_{15\times7})$. By augmenting these 
two matrices, we can construct ${\bf A}$. The matrix $B^1$ is given as:
\vspace{0.5cm}
\begin{equation}
\small
 \begin{pmatrix}
  2F'_{11} & 2F'_{12}		& 	2F'_{13}		&	   2F'_{14}	&	  2F'_{15}	&	0	&		0		&	0\\
  F'_{21}  & (F'_{11}+F'_{22})& F'_{23}			&	    F'_{24}	&	   F'_{25}	& F'_{12} &	F'_{13} 		& F'_{14}\\
  F'_{31}  & F'_{32}		& (F'_{11}+F'_{33}) &	    F'_{34}	&	   F'_{35}	&	0	&	F'_{12}		&	0\\
  F'_{41}  & F'_{42}		& F'_{43}			& (F'_{11}+F'_{44}) &	   F'_{45}	&	0	&		0		& F'_{12}\\
  F'_{51}  & F'_{52}		& F'_{53}			&	    F'_{54}	& (F'_{11}+F'_{55}) &	0	&		0		&	0\\
        0        & 2F'_{21}		& 		0		&		0		&		0		& 2F'_{22}& 	2F'_{23}		& 2F'_{24}\\
        0	 & F'_{31}		& 	F'_{21}		&		0		&		0		& F'_{32} & (F'_{22}+F'_{33}) & F'_{34}\\
        0	 & F'_{41}		&		0		&	    F'_{21}	&		0		& F'_{42} &	 F'_{43}		& (F'_{22}+F'_{44})\\
        0        &    F'_{51}	& 	    0		&       0	& F'_{21} & F'_{52} & F'_{53} & F'_{54} \\
        0        &          0	 	&    2 F'_{31}	&       0	&	0	&	0	&2F'_{32}	&	0\\
        0        &          0	 	&     F'_{41}	& F'_{31} &	0	&	0	& F'_{42} & F'_{32}\\
        0        &          0	 	&     F'_{51}	&       0	& F'_{31} &	0	& F'_{52} &	0\\
        0        &          0	 	& 	    0		&2F'_{41} &	0	&	0	&	0	&2F'_{42}\\
        0	 &          0	 	& 	    0		& F'_{51} & F'_{41} &	0	&	0	& F'_{52}\\
        0	 &          0	 	& 	    0		&       0	&2F'_{51} &	0	&	0	&	0
 \end{pmatrix}
\end{equation}

Similarly, the matrix $B^2$ is given as:
\begin{equation}
\small
 \begin{pmatrix}
        0        &		0		& 		0		&		0		&		0		&	0	&	0	\\
  F'_{15}  &		0		& 		0		&		0		&		0		&	0	&	0	\\
        0        &	 F'_{13}		&	   F'_{14}	&	   F'_{15}	&		0		&	0	&	0	\\
        0        &		0		&	   F'_{13}	&		0		&	  F'_{14}		& F'_{15} &	0	\\
  F'_{12}  &		0		&		0		&	   F'_{13}	&		0		& F'_{14} & F'_{15} \\
 2F'_{25}  &		0		& 		0		&		0		&		0		&	0	&	0	\\
  F'_{35}  &	 F'_{23}		& 	 F'_{24}		&	   F'_{25}	&		0		&	0	&	0	\\
  F'_{45}  &		0		&	 F'_{23}		&		0		&	  F'_{24}		& F'_{25} &	0	\\
 (F'_{22}+F'_{55})	&		0		&		0		&	   F'_{23}	&		0		&	  F'_{24}		&	  F'_{25}	\\
		0		&	2F'_{33}		&	2F'_{34}		&	  2F'_{35}	&		0		&		0		&		0	\\
		0		&	 F'_{43}		& (F'_{33}+F'_{44})	&	   F'_{45}	&	   F'_{34}	&	  F'_{35}		&		0	\\
	  F'_{32}		&	 F'_{53}		&	 F'_{54}		& (F'_{33}+F'_{55})	&		0		&	  F'_{34}		&	  F'_{35}	\\
		0		&		0		&	 2F'_{43}		&		0		&	 2F'_{44}		&	 2F'_{45}		&		0	\\
	  F'_{42}		&		0		&	  F'_{53}		&	  F'_{43}		&	  F'_{54}		& (F'_{44}+F'_{55})	&	  F'_{45}	\\
	 2F'_{52}		&		0		&		0		&	 2F'_{53}		&		0		&	 2F'_{54}		&	 2F'_{55}
 \end{pmatrix}
\end{equation}

\section{}\label{AppendB}
A typical case of fitting an up-going muon track of momentum 2 GeV/c at initial direction $\theta=
\cos^{-1}0.75$ is considered. 
In a small step inside the magnetized iron plate, we compare the coefficients of the powers of $l$, 
$l^2$ and $l^3$ to check the consistency between Eq.~\eqref{MankelQ} and Eq.~\eqref{Eq15d}.

\begin{table}[ht]
\small
\hspace{-1.0cm}                         
\begin{tabular}{|c||c|c|c||c|c|c||}
\hline
{Process noise} & \multicolumn{3}{ c|| }{Mankel [Eq.~\eqref{MankelQ}], $p_{rec}=2.258$ GeV/c} & \multicolumn{3}{ c ||}{Modified [Eq.~\eqref{Eq15d}], $p_{rec}=2.259$ GeV/c}\\\cline{1-7}
\hline
 Element & coeff($l$) & coeff($l^2$) & coeff($l^3$) & coeff($l$) & coeff($l^2$) & coeff($l^3$)    \\\hline
$Q_{11}$ &     0      &       0      &   0.037123   & $\sim10^{-17}$ & $\sim10^{-17}$ & 0.037129  \\\hline
$Q_{12}$ &     0      &       0      &   0.001492   & $\sim10^{-17}$ & $\sim10^{-17}$ & 0.001478  \\\hline
$Q_{13}$ &     0      &  0.055684    &       0      & $\sim10^{-18}$ & 0.055689       & 0.010299  \\\hline
$Q_{14}$ &     0      &  0.002239    &       0      & $\sim10^{-18}$ & 0.002228       &{\bf -0.020923}  \\\hline
$Q_{22}$ &     0      &       0      &   0.021937   & $\sim10^{-16}$ & $\sim10^{-17}$ & 0.021935  \\\hline
$Q_{23}$ &     0      &  0.002239    &       0      & $\sim10^{-16}$ & 0.002228       &{\bf -0.020915}  \\\hline
$Q_{24}$ &     0      &  0.032906    &       0      & $\sim10^{-17}$ & 0.032905       &-0.002956  \\\hline
$Q_{33}$ &  0.111369  &       0      &       0      & 0.111369       & 0.020610       &-0.034827  \\\hline
$Q_{34}$ &  0.004477  &       0      &       0      & 0.004477       &-0.041840       & 0.007156  \\\hline
$Q_{44}$ &  0.065812  &       0      &       0      & 0.065812       &-0.005924       & 0.014649  \\\hline
\end{tabular}
\caption{Comparison of coefficients of $l$, $l^2$ and $l^3$ inside iron plate of ICAL (within magnetic field)
to test the validity of Eq.\eqref{Eq15d}
\label{Table}}                     
\end{table}
It is seen from Table~\ref{Table} 
that the corresponding coefficients are close enough. For example, the element $Q_{24}=0.032906\ l^2
$ when calculated from Eq.~\eqref{MankelQ}. In the modified approach, the value of $Q_{24}$ becomes: 
$Q_{24}=(10^{-17})\ l+0.032905\ l^2-0.002956\ l^3$ where $l\sim10^{-3}\rm{ m}$. So, the modification 
introduces very weak correction. In this case, the reconstructed momentum with the new process noise 
matrix became $|p_{rec}|=2.259$ GeV/c compared to $|p_{rec}|=2.258$ GeV/c, estimated by the standard 
approach. The table also shows the interesting point that the difference between $Q_{14}$ and $Q_{23
}$ happens at the order of $l^3$. In Eq.~\eqref{MankelQ}, these two $independent$ elements were the same! 

\section{}\label{AppendC}
For computations in the high energy physics experiments, ROOT~\cite{Brun:1997pa} is a widely accepted 
software. However, because of the inevitable occurrence of the complex numbers in this problem, it is 
rather difficult to implement the recipe of Eq.\eqref{Eq15d} using ROOT. The reason is following: the 
actual diagonal matrix becomes block-diagonal in the convention followed by ROOT, since it pushes the 
imaginary parts of the eigenvalues to the off-diagonal positions (see: `Matrix Eigen Analysis' in 
Chapter 14 of~\cite{brun2003root}). The eigenvector matrix is also kept real in ROOT. 

However, we wanted to proceed with the standard diagonalization method for 
which all the eigenvalues, real or complex, appear at the diagonal position. Therefore, we used a C++ 
based library it++~\cite{cristeait++}. This library can be easily interfaced with existing code which 
is written in C++ by appending `{\tt itpp-config --cflags}' and `{\tt itpp-config --libs}' to LDFLAGS 
in the GNUMakefile. This library can be easily used to find eigenvalues and eigenvectors of ${\bf A}$ 
in the standard forms. The following it++ member function was used: {\tt itpp::eig(${\bf const\ mat}\ 
\&{A},{\bf cvec}\ \&d,{\bf cmat}\ \&P$)} to carry out the procedure. In this function, 
$d$ denotes the $\underline{c}$omplex vector of eigenvalues and $P$ denotes the $\underline{c}$omplex 
matrix obtained by augmenting the eigenvectors of ${\bf{A}}$. This matrix is seen to have determinant 
nonzero and thus, is invertible. As required by Eq.\eqref{Eq15d}, the inverse matrix $P^{-1}$ is made 
to operate on $\delta{\bf{q}}$ and further computations are performed.

This package is based on external computational libraries, like {\tt BLAS}~\cite{lawson1979basic} and 
{\tt LAPACK}~\cite{anderson1990lapack}. The level of accuracy of the computation is seen to be of the 
same order as of Mathematica~\cite{wolfram1999mathematica}. For example, the eigenvalues and eigenvectors 
of a matrix computed by it++ and Mathematica are found to be consistent within $\sim 1\%$. 

\section*{References}
\bibliographystyle{unsrt}
\biboptions{sort&compress}
\bibliography{KFpNoise}

\end{document}